\documentclass[12pt,a4paper]{article}
\usepackage{amsmath,amssymb}
\usepackage{graphicx}
\usepackage{cite}

\allowdisplaybreaks[3]

\begin{document}

\begin{titlepage}

\begin{flushright}
KEK-TH-2102, NCTS-TH/1902
\end{flushright}

\vspace{5em}

\begin{center}
{\Large\textbf{Entropy Generation at the Horizon Diffuses Cosmological Constant in 2D de Sitter Space}}
\end{center}

\begin{center}
Hiroyuki \textsc{Kitamoto}$^{1)}$
\footnote{E-mail address: kitamoto@cts.nthu.edu.tw}
Yoshihisa \textsc{Kitazawa}$^{2),3)}$
\footnote{E-mail address: kitazawa@post.kek.jp} 

\end{center}

\begin{center}
$^{1)}$
\textit{Physics Division, National Center for Theoretical Sciences}\\
\textit{National Tsing-Hua University, Hsinchu 30013, Taiwan}\\
$^{2)}$
{\it KEK Theory Center, Tsukuba, Ibaraki 305-0801, Japan}\\
$^{3)}$
{\it Department of Particle and Nuclear Physics}\\
{\it The Graduate University for Advanced Studies (Sokendai)}\\
{\it Tsukuba, Ibaraki 305-0801, Japan}
\end{center}

\begin{abstract}
We investigate a solution of the exactly renormalized Liouville action 
to foresee the fate of the two-dimensional de Sitter space. 
We work in the semiclassical region with a large matter central charge $c$. 
Instead of de Sitter expansion, it performs a slow-roll inflation with the parameters $\epsilon=(1/2)\eta =6/c$. 
An inflaton field is induced in the effective theory to describe quantum effects of the Liouville theory. 
The geometric entropy increases logarithmically with the Hubble radius. 
We propose that de Sitter entropy is carried by superhorizon modes of the metric. 
It can be directly estimated from the partition function as $S=\log Z$ in Liouville gravity. 
We formulate a gravitational Fokker-Planck equation to elucidate the Brownian process at the horizon: 
the superhorizon modes are constantly jolted by newcomers. 
We show that such a built-in entropy-generating process diffuses the cosmological constant. 
We evaluate von Neumann entropy associated with the distribution function of superhorizon modes.
It always increases under the Fokker-Planck equation in a consistent way with semiclassical estimates.
The maximum entropy principle operates in quantum gravity. 
An analogous entropy production mechanism at the horizon might have increased the Hubble radius 
much beyond the microscopic physics scale in the Universe. 
\end{abstract}

\vspace{\fill}

May 2019

\end{titlepage}

\section{Introduction}
\setcounter{equation}{0}

It is often suspected that the cosmological constant problem is an infrared problem. 
De Sitter space is unstable due to some shielding effects or particle productions \cite{Polyakov}. 
In particular, the importance of IR logarithmic effects is stressed 
to break de Sitter symmetry \cite{Woodard1996}. 
A stochastic approach is proposed to sum up leading IR logarithms \cite{Starobinsky,Woodard2005}. 
In order to understand physics in four-dimensional de Sitter space, 
two-dimensional exactly solvable models may be instructive. 
In this paper, we investigate two-dimensional de Sitter space in Liouville quantum gravity. 
It is realized as a classical solution in the semiclassical regime with a large matter central charge. 
By examining the exact solution, 
we note that the negative anomalous dimension of the cosmological constant operator makes 
the Hubble parameter time dependent and it fades away. 
The shielding effect of the cosmological constant arises
due to the negative sign of the kinetic term of the conformal mode. 
IR logarithmic effects are important to make the Hubble parameter time dependent.  
We estimate semiclassical geometric entropy of the two-dimensional de Sitter space. 
It grows logarithmically with the inverse Hubble parameter $H$ like $S\sim \log(1/H)$. 
We propose that it is carried by the superhorizon mode of the conformal degrees of the metric. 
We investigate its distribution function by Fokker-Planck equations. 
We offer evidences that the von Neumann entropy of the distribution function can be identified 
with the geometric entropy of the space. 

The presence of event horizons leads us to the thermodynamics of black holes \cite{Bousso}. 
Gibbons and Hawking showed that analogous relations hold in de Sitter space 
with cosmological horizons \cite{Gibbons}. 
The area of the cosmological horizon is 
\begin{align}
A_0=4\pi l^2,
\end{align}
where the surface gravity is $\kappa=1/l$.
A short summary of de Sitter-space thermodynamics is given in Appendix A.

The metric of de Sitter space in global coordinates is
\begin{align}
\frac{ds^2}{l^2}=-d\tau^2+\cosh^2(\tau) d\Omega_3^2, 
\label{GDS1}\end{align} 
while it becomes conformally flat in the local Poincar\'e patch: 
\begin{align}
ds^2=-dt^2+e^{2Ht}(dr^2+r^2d\Omega_2^2)
=\big(\frac{1}{-H\tau}\big)^2(-d\tau^2+d\vec{x}^2). 
\label{LPC1}\end{align} 
The metric is periodic under the shift of time into the imaginary direction by $2\pi l$. 
The Green's functions must have the identical periodicity.
This implies that the temperature of the cosmological horizon is
\begin{align}
T_\text{dS}=\frac{1}{2\pi l}. 
\end{align}
We may rotate the de Sitter space in the global coordinates (\ref{GDS1}) 
into Euclidean $S^4$ with the metric
\begin{align}
\frac{ds^2}{l^2}=d\chi^2+\sin^2(\chi) d\Omega_3^2.
\end{align}
The Euclidean action on $S^4$ with the radius $1/H$ is
\begin{align}
\int d^4x \sqrt{g}\frac{1}{8\pi G_N}\big(\frac{1}{2}R - 3H^2\big)
=\int d^4x \sqrt{g}\frac{3H^2}{8\pi G_N} 
=\frac{8\pi^2}{3H^4}\frac{3H^2}{8\pi G_N}. 
\end{align}
The classical action gives us the entropy of de Sitter space once we put  the solutions in it. 
In quantum gravity, the size of the space is a dynamical variable to be integrated over. 
In general, the partition function $Z$ must be stationary with respect to the change of size $l$ 
or the inverse temperature $\beta=2\pi l$ of the manifold: 
\begin{align}
\frac{\partial}{\partial \beta}\log Z =0\ \Rightarrow\ 
S=\big(1-\beta\frac{\partial}{\partial \beta}\big)\log Z =\log Z.
\end{align}
At the semiclassical level, the entropy of de Sitter space is given by
\begin{align}
S_0=\frac{A_0}{4G_N}=\frac{\pi}{H^2G_N}. 
\label{4DEP}\end{align}
The Euclidean quantum gravity represents a possible equilibrium state with the temperature set 
by the scale of the event horizon. 
It may have instability since the Einstein action is not bounded below. 
The kinetic term of the scale factor of the metric (conformal mode) is of the wrong sign. 
There is no fundamental remedy for this problem and we mostly work with Lorentz signature. 
The situation is better with respect to the superhorizon modes as the potential term dominates
over the kinetic term. 
The thermodynamics of superhorizon modes may be studied in Euclidean gravity. 

Let us start our investigation on 2D quantum gravity which has de Sitter space 
as a classical solution at the tree level. 
We focus on the dynamics of the conformal degrees of the metric $\phi$
by choosing the conformal gauge as follows: 
\begin{align}
g_{\mu\nu}=e^{\phi}\hat{g}_{\mu\nu}. 
\end{align}
In this expression, $\hat{g}_{\mu\nu}$ denotes the classical background of the metric.
In 2D quantum gravity, we work with the Liouville action which is the gift of the conformal anomaly\footnote{
A short summary of the derivation of  the Liouville action is given in Appendix A.}: 
\begin{align}
\frac{Q^2}{4\pi}\int d^2x\sqrt{\hat{g}}
\big(\frac{1}{4}\hat{g}^{\mu\nu} \partial_{\mu}\phi \partial_{\nu}\phi 
+\frac{1}{2}\phi \hat{R}-H^2e^{\phi}\big).
\end{align}
The equation of motion with respect to constant $\phi$ is
\begin{align}
\hat{R}=2 H^2e^{\phi}. 
\label{CLEQ}\end{align}
The two-dimensional symmetric space $S^2$ is the solution with the scale  $e^{\phi}=1/H^2$.
By rotating two-dimensional de Sitter space into $S^2$, 
we obtain a semiclassical estimate of the geometric entropy, 
\begin{align}
\frac{Q^2}{8\pi}\int d^2x\sqrt{\hat{g}}(\phi \hat{R}-\hat{R})={Q^2}\log \big(\frac{1}{H^2}\big). 
\label{CLGE}\end{align}
This should be contrasted with (\ref{4DEP}) in four dimensions. 
We note that the inverse Newton's coupling constant is replaced by $Q^2=(c-25)/6$ in two dimensions.\footnote{There is a quantum correction to this estimate due to the anomalous dimension of
the cosmological constant  operator. $Q^2$ should be replaced by $Q^2/\gamma$.}
The matter central charge $c$ must be larger than 25 to correspond with a positive Newton's coupling. 
It implies that the sign of the kinetic term of the conformal mode is negative. 
We work in this semiclassical region.
It also depends on the Hubble parameter $H^2$. 
Although the entropy increases if $H^2$ decreases in both cases, 
its dependence is weak in two dimensions: $\log (1/H^2)$, while much stronger in four dimensions: $1/H^2$. 

Nevertheless, this toy model may reveal to us what carries the entropy of de Sitter space \cite{NI,NCHUS}.
Furthermore, we identify a mechanism to create entropy 
as more and more degrees of freedom are going out of the horizon. 
Simultaneously, the cosmological constant is diffused away. 
Since we find the common prerequisite between two- and four-dimensional de Sitter spaces 
for such a mechanism to work, 
the cosmological constant $\Lambda$ may have decreased
far beyond the scale of the microscopic physics in an analogous mechanism. 
Certainly, explaining why the dimensionless ratio $H^2G_N\sim 10^{-120}$ is so small is a difficult problem. 
However, the most important step in solving the problem is to correctly formulate it. 
It may be a more appropriate question to ask why the current Universe has such a huge entropy. 
As the above equation (\ref{4DEP}) shows, the entropy of four-dimensional de Sitter space 
is inversely proportional to the Hubble parameter $1/H^2$. 
Since the entropy always increases, the Hubble parameter must decrease 
especially when there are huge entropy creations in the Universe \cite{Frolov}. 
Such a simple trend seems to capture the essence of the history of the Universe. 

This concludes our introductory section to this paper. 
In Section 2, we investigate a solution of exactly renormalized Liouville action. 
Two-dimensional de Sitter space appears as a solution of nonrenormalized Liouville action. 
Fortunately, the effects of short-distance divergences are known exactly. 
In two dimensions, UV and IR divergences are closely related 
since the propagators do not change unlike in four dimensions.
We can thus predict IR behavior from the UV behavior. 
In the semiclassical region with a large matter central charge $c$, 
the scaling dimension $\gamma$ of the cosmological constant operator is less than the canonical $\gamma < 1$. 
In such a case, we show that de Sitter expansion becomes slow-roll inflation with the slow-roll parameters 
inversely proportional to $c$. 
In Section 3, we propose that de Sitter entropy is the von Neumann entropy of superhorizon modes.
We show that the entropy is generated at the rate $\dot{S}=2H$ by the stochastic equations 
which diffuses the cosmological constant. 
We conclude in Section 4. 
Basic information are summarized in three appendices for self-containedness: 
Appendix A on de Sitter thermodynamics, 
Appendix B on operator renormalization in Liouville theory 
and Appendix D on stochastic equations. 
In Appendix C, we point out a dual description 
for a class of two-dimensional quantum gravity models. 

\section{Solutions of  2D de Sitter quantum gravity}\label{S-Liouville}
\setcounter{equation}{0}

In this section, we investigate the quantum IR effects in an exactly solvable model: 
two-dimensional quantum gravity. 
We adopt a conformal gauge and parametrize the metric as 
\begin{align}
g_{\mu\nu}=e^{\phi}\hat{g}_{\mu\nu}, 
\end{align}
where $\hat{g}_{\mu\nu}$ is a background metric. 
The metric is Lorenzian corresponding to real spacetime. 
The effective action for the conformal mode $\phi$ is the Liouville action:
\begin{align}
\int \sqrt{-\hat{g}}d^2x \big\{\frac{c-25}{96\pi}(\hat{g}^{\mu\nu}\partial_{\mu}\phi\partial_{\nu}\phi
+2\phi\hat{R})-\Lambda e^{\gamma\phi}\big\}. 
\label{Liouville}\end{align}
Here $c$ denotes the central charge of the matter minimally coupled to two-dimensional quantum gravity.
In the free-field case, $c$ counts massless scalars and fermionic fields as $c=N_s+N_f/2$.
We consider the semiclassical regime: $c > 25$, 
where the sign of the kinetic term for the conformal mode is negative and hence timelike. 
The identical feature occurs with four-dimensional Einstein gravity. 
In the above expression, $\Lambda$ is the cosmological constant, 
$e^{\gamma\phi}$ is a renormalized cosmological constant operator 
and $\gamma-1$ denotes the anomalous dimension. 

The equation of motion with respect to $\phi$ is given by 
\begin{align}
\frac{Q^2}{8\pi}\nabla^2\phi+\Lambda e^{\phi}=0, 
\end{align}
where we set the background scalar curvature $\hat{R}=0$ anticipating to adopt conformally flat coordinates.
$Q^2$ is the effective inverse Newton's coupling in two dimensions and we also set $\gamma \rightarrow 1$.
This is allowed in the large-$c$ limit and the classical geometry holds. 
In contrast, geometry is quantized when $\gamma < 1$. 

Furthermore, there is another equation of motion 
with respect to the traceless mode of the metric $h^{\mu\nu}$: 
\begin{align}
\frac{Q^2}{8\pi}\big\{(\nabla_\mu\phi\nabla_\nu\phi-2\nabla_\mu\nabla_\nu\phi)
-\frac{1}{2}\hat{g}_{\mu\nu}(\nabla_\rho\phi\nabla^\rho\phi-2\nabla^2\phi)\big\}
=\nabla_\mu \chi\nabla_\nu \chi-\frac{1}{2}\hat{g}_{\mu\nu}\nabla_\rho \chi\nabla^\rho \chi, 
\label{HEQ}\end{align}
where $\chi$ denotes a free scalar field 
and the presence of the $\phi\hat{R}$ term in the Liouville action results 
in the linear term in $\phi$ on the left-hand side. 

For the conformally flat metric $g_{\mu\nu}=e^{\phi}\eta_{\mu\nu}$,
the action (\ref{Liouville}) is simplified as 
\begin{align}
\int d^2x \big\{\frac{Q^2}{16\pi}(\eta^{\mu\nu}\partial_\mu\phi\partial_\nu\phi-4H^2 e^{\phi})\big\}.
\label{LFG}\end{align}
The effective Newton's coupling constant $G_N$ becomes small when $c$ becomes large like $1/Q^2\sim {6/c}$. 
We confine our investigation of this model to the semiclassical region $c>25$. 
The kinetic term of the conformal made $\phi$ is negative as we can see in (\ref{LFG}). 
This feature is shared with four-dimensional Einstein gravity. 
The expansion of the Universe occurs due to such an instability. 
For theorists, the wrong sign of the  kinetic term of $\phi$ looks like a curse 
against going beyond semiclassical investigations. 
On the contrary, it could be a blessing 
as we demonstrate by investigating a two-dimensional toy model in this paper. 
It provides a mechanism to grow the Universe to be vast in comparison to the microscopic physics scale. 
Namely, the cosmological constant is shielded by quantum fluctuations of the conformal mode $\phi$. 
The shielding occurs as more entropy is generated at the cosmological horizon 
as the conformal zero mode $\phi_0$ accumulates at the horizon. 
The fluctuations perform Brownian motion as the superhorizon modes are constantly jolted by newcomers. 
The negative kinetic term of $\phi$ is crucial for shielding the cosmological constant. 
Note that the Hubble parameter $H^2$ is reduced by $Q^2$ when we fix the cosmological constant $\Lambda$: 
\begin{align}
H^2=G_N \Lambda = \frac{4\pi}{Q^2}\Lambda. 
\end{align}
With large $c$, matter fields reduce $H^2$ by a large factor as $H^2\sim \frac{24\pi}{c}\Lambda$.

The classical solution of this action (\ref{LFG}) is two-dimensional de Sitter (dS) space with the metric 
\begin{align}
\phi_c=-2\log (-H\tau)=\log a^2(t),\hspace{1em}
e^{\phi_c}=\big(\frac{1}{-H\tau}\big)^2 = e^{2Ht}. 
\label{URSL}\end{align}
By adopting the Poincar\'e coordinate, we obtain the following metric 
as a solution of the classical Liouville theory:  
\begin{align}
ds^2=\big(\frac{1}{-H\tau}\big)^2(-d\tau^2 +dx^2)=-dt^2+e^{2Ht}dx^2.
\label{metric}\end{align}
It covers the upper half-triangle of the Penrose diagram of the global de Sitter manifold. 
The volume operator is of the identical form with the metric in accordance with the classical geometry. 
We identify cosmic time with the  classical solution of the conformal mode $\phi_c (t)=2H t$. 
The distance of the cosmic horizon from the observer is $1/H$. 
She or he is situated at the center of the line segment. 
On the other hand, if the cosmological constant can be neglected, 
we also have a solution with a nontrivial free matter field $\chi$: 
\begin{align}
\phi_c = A\tau,\hspace{1em}\chi_c=A\sqrt{\frac{Q^2}{8\pi}}\tau ,
\end{align}
where $A$ is an arbitrary constant. 
This is a two-dimensional Friedmann spacetime. 
This solution should go over to the two-dimensional dS-space solution 
when the cosmological constant becomes dominant. 
Another solution is obtained by adding the gas of massless scalar particles with temperature $T$ 
to empty de Sitter space. 
We may solve (\ref{HEQ}) to linear order in perturbation $\phi_c+\phi$ as 
\begin{align}
\frac{Q^2}{4\pi}\left(\nabla_{0}\phi_c\nabla_{0}\phi-\nabla_{0}\nabla_{0}\phi\right)
=2\langle\nabla_{0}\chi\nabla_{0}\chi\rangle. 
\end{align}
Specifically, 
\begin{align}
\frac{Q^2}{4\pi}\big(\frac{2}{-\tau}\nabla_{0}\phi-\nabla_{0}\nabla_{0}\phi\big)=\frac{\pi}{3}{T^2}
\ \Rightarrow\ \phi=-\frac{\pi^2}{15 Q^2}\frac{T^2}{H^2 a^2(t)}. 
\end{align}
Since their energy density decays like $T^2/a^2(t)$ with the expansion of the Universe, 
the contribution of the massless scalar particles fades away 
in comparison to the cosmological constant at a late time. 
However, their contribution to entropy remains constant as can be seen from (\ref{CLGE}).\footnote{
This is essentially the entangled entropy. 
There are important quantum corrections to this classical formula, which is the subject of this work.}
They contribute $c=1$ to the coefficient of the Liouville Lagrangian. 
In this way, the massless fields leave their legacy in reducing the Hubble parameter. 

We expand the action around the classical background $\phi_c+\phi$:
\begin{align}
&\int d^2x  \frac{Q^2}{8\pi}\big\{-\frac{1}{2}(1+h^{00})
\frac{\partial}{\partial\tau}\phi_c\frac{\partial}{\partial\tau}\phi_c 
+h^{00}\frac{\partial ^2}{\partial\tau^2}\phi_c -2H^2 e^{\phi_c}\big\} \notag\\
+&\int d^2x \frac{Q^2}{8\pi}\sqrt{-\hat{g}}\big\{\frac{1}{2}\hat{g}^{\mu\nu}
\frac{\partial}{\partial x^\mu}\phi\frac{\partial}{\partial x^\nu}\phi
+(1+\phi)\hat{R}-{2H^2}e^\phi\big\}, 
\end{align}
where 
\begin{align}
\sqrt{-\hat{g}}=e^{\phi_c},\hspace{1em}
\sqrt{-\hat{g}}\hat{R}=\frac{\partial^2}{\partial \tau^2}\phi_c=2H^2\sqrt{-\hat{g}}. 
\label{DFGQ}\end{align}
Note that we have recovered the Liouville action on the de Sitter space (\ref{Liouville}) 
for quantum fluctuations. 
We then renormalize the cosmological constant operator as we review it briefly in Appendix B.
The results of the investigations from various viewpoints agree 
that the cosmological constant operator is renormalized as
\begin{align}
e^{\phi}\rightarrow e^{\gamma\phi}.
\end{align}
At the short-distance limit, the volume operator is of the original form $e^\phi$.
It is renormalized  to become $e^{\gamma\phi}$ in the IR limit under the
renormalization group equation (\ref{DFEQ}). It is a diffusion equation which plays a crucial role in this work. 
Both the conformal invariance and renormalization group arguments show that 
the scaling dimension satisfies the following relation 
\begin{align}
\gamma+ \frac{\gamma^2}{Q^2}=1.
\label{saddle}\end{align}
By solving this equation, the scaling dimension of the cosmological constant operator is determined to all orders 
\begin{align}
\gamma=\frac{2}{1+\sqrt{1+\frac{4}{Q^2}}}= 1-\frac{1}{Q^2} + 2\big(\frac{1}{Q^2}\big)^2+\cdots. 
\label{anmd}\end{align}
The important feature is that $\gamma < 1$; 
namely, it is smaller than the canonical dimension in the semiclassical region.

Since the cosmological constant operator changes after the renormalization, 
we may look for a new classical solution of the following type of action: 
\begin{align}
\int d^2x \big\{\frac{Q^2}{16\pi}(\eta^{\mu\nu}\partial_{\mu}\phi\partial_{\nu}\phi
-\frac{4H^2}{\gamma} e^{\gamma\phi})\big\}.
\label{LFGN}\end{align}
It is apparent from the above action in comparison to nonrenormalized one (\ref{LFG})
that a reinterpretation of $\phi$ as $\gamma\phi $ may give us a solution of the renormalized action (\ref{LFGN}). 

However, such a candidate cannot satisfy the equation of motion with respect to $h^{00}$ (\ref{HEQ}) 
since it is nonlinear and does not allow a simple scaling transformation. 
In order to circumvent this problem, we introduce an inflaton field $f$ in such a way that (\ref{HEQ}) is satisfied. 
We argue that it is a standard strategy to enlarge field spaces in order to solve the highly nonlinear problems. 

Let us postulate the following Lagrangian: 
\begin{align}
\int d^2 x \frac{Q^2}{8\pi}\sqrt{-\hat{g}}\big\{-\frac{1-\gamma}{2}\hat{g}^{\mu\nu}\partial_\mu f \partial_\nu f
+\frac{1}{2}(\hat{g}^{\mu\nu}\partial_\mu \phi \partial_\nu \phi +2 \hat{R}\phi)
-\frac{2H^2}{\gamma}e^{\phi-(1-\gamma)f}\big\}. 
\label{IFAT}\end{align}
The equations of motion with respect to the inflaton $f$ and the conformal mode $\phi$ are
\begin{align}
\nabla_0^2\gamma f = 2H^2e^{\phi-(1-\gamma)f},\hspace{1em}
\nabla_0^2 \gamma \phi = {2H^2} e^{\phi-(1-\gamma)f}, 
\label{IFTS}\end{align}
where we assume $\hat{R}=0$ in a conformally flat gauge. 
We can identify $f=\phi$ and the inflaton field $f$ adds the following term on the right-hand side of (\ref{HEQ}): 
\begin{align}
\frac{Q^2}{8\pi}(1-\gamma)
\big\{\nabla_\mu f\nabla_\nu f-\frac{1}{2}\hat{g}_{\mu\nu}\nabla_\rho f\nabla^\rho f\big\}. 
\end{align}
Then both sides of (\ref{HEQ}) coincide as a result of the introduction of the inflaton.   
Furthermore, the cosmological constant operator becomes
\begin{align}
e^{\phi-(1-\gamma)f}\rightarrow e^{\gamma\phi}=\big(\frac{1}{-H \tau}\big)^2. 
\end{align}
This is a $(1,1)$-type operator, which is consistent with conformal invariance. 
In this way, we obtain a solution which satisfies all equations of motion and required symmetries. 

After putting the inflaton field under the rug with the identification $f=\phi$, we obtain the following action\begin{align}
\int d^2 x \frac{Q^2}{8\pi}\sqrt{-\hat{g}}
\big\{\frac{\gamma}{2} \hat{g}^{\mu\nu}\partial_\mu \phi \partial_\nu \phi 
+\hat{R}\phi-\frac{2H^2}{\gamma}e^{\gamma \phi}\big\}.
\label{ACDU}\end{align}
We regard this theory as an effective field theory equivalent to the fully quantized Liouville theory.  
A solution of 2D quantum gravity is given by a slow-roll inflation. 
We should be careful to avoid double counting  when we elucidate the predictions of this theory. 
For example, the quantum effects of $\phi$ on the Hubble parameter are taken into account 
by the classical motion of the inflaton $f$ (\ref{HBWB1}) at the weak-coupling or slow-roll limit: 
\begin{align}
H^2(t)=H^2e^{-(1-\gamma)f(t)}\sim H^2\exp{(-\frac{1}{Q^2}2Ht)}.
\end{align}

After rescaling the fields $\gamma\phi\rightarrow \phi$, our action takes the following form: 
\begin{align}
\int d^2 x  \frac{Q_I^2}{8\pi}
\big(-\frac{1}{2}\frac{\partial}{\partial\tau}\phi_c\frac{\partial}{\partial\tau}\phi_c -2H^2e^{\phi_c}\big)
\label{CLLV}\end{align}
\begin{align}
+\int d^2x \frac{Q_I^2}{8\pi}\sqrt{-\hat{g}}
\big\{\frac{1}{2}\hat{g}^{\mu\nu}\frac{\partial}{\partial x^\mu}\phi\frac{\partial }{\partial x^\nu}\phi
+\phi \hat{R}-{2H^2}(e^{\phi}-1)\big\}.
\label{CQSF}\end{align}
This is nothing but the nonrenormalized action and its solution is the original one (\ref{URSL}). 
The important difference is that $Q_I^2=Q^2/\gamma$ appears 
in the effective inverse Newton's coupling $G_N$. 
In the end, the solution of the renormalized action can be obtained from the nonrenormalized one 
by a simple scaling $\phi_c\rightarrow \gamma\phi_c$. 
The geometric objects are defined by new solutions after $\phi_c$ is reinterpreted as $\gamma \phi_c$. 
We confirm that this action explains the scaling relations between the Newton's coupling $G_N$ 
and the Hubble parameter $H^2$ correctly: 
\begin{align}
\sqrt{-\hat{g}}=e^{\gamma\phi_c},\hspace{1em}
\sqrt{-\hat{g}}\hat{R}=\frac{\partial^2}{\partial \tau^2}\gamma\phi_c=2H^2\sqrt{-\hat{g}}. 
\label{DFGQ1}\end{align}
It is important to recognize that the effective inverse Newton's coupling $Q^2$ is replaced by $Q^2/\gamma$. 
This fact implies that the physical scale has changed by the factor $\gamma$. 
We thus believe that this recycling of an old solution as a new one is not vacuous, but a scale transformation. 

We can read off the scaling relations between the Hubble parameter and the topological coupling $G_T$ 
in front of the scalar curvature $\phi \hat{R}$ term as we explain below. 
It is identical to Newton's coupling $\ G_T=G_N$. 
It has been useful to consider the response of the action 
under $\phi_c\rightarrow \phi_c - \varphi$ and $\phi\rightarrow \phi +\varphi$. 
Of course, the action is invariant if $\varphi$ is a local conformal transformation since we start with (\ref{LFG}). In fact, conformal invariance has been an effective tool 
to determine the renormalized Liouville action \cite{DK,FD}. 

The action (\ref{CQSF}) implies that the semiclassical entropy of two-dimensional de Sitter space is 
\begin{align}
S_c=\frac{Q^2}{\gamma}\varphi=\frac{Q^2}{\gamma}\log \frac{H_0^2}{H^2(t)}
=\frac{1}{G_N}\log \frac{H_0^2}{H^2(t)}, 
\label{SCEP}\end{align}
where $H_0$ denotes an initial value of the Hubble parameter. 
The renormalization of the cosmological constant operator has introduced $\gamma$ dependence.
It depends not only on the gravitational coupling but also on the Hubble parameter.
We can predict the relative scaling relation between the Hubble parameter  $H^2$ 
and the effective Newton's coupling from (\ref{CQSF}). 
Let us assume that the Hubble parameter changes slowly over cosmic time evolution. 
We may postulate 
\begin{align}
H^2(t)=H_0^2e^{-\varphi}.
\end{align}
We consider a constant shift of the quantum field
\begin{align}
\phi(x) \rightarrow \phi(x)+\varphi, 
\end{align}
\begin{align}
\frac{Q^2}{8\pi\gamma}\int d^2 x \sqrt{-\hat{g}}(\phi\hat{R}-2H^2(t)e^{\phi})\ \rightarrow\ 
\frac{Q^2}{8\pi\gamma}\int d^2 x\sqrt{-\hat{g}} \{(\phi +\varphi) \hat{R}-2H_0^2e^{\phi }\}.
\end{align}
The action changes as
\begin{align}
i \frac{Q^2}{8\pi\gamma}\int_{dS^2} d^2 x\sqrt{-\hat{g}} \varphi \hat{R}\ 
\rightarrow\ \frac{Q^2}{8\pi\gamma}\int_{S^2} d^2 x\sqrt{\hat{g}} \varphi \hat{R}=\frac{Q^2}{\gamma} \varphi .
\end{align}
In the last step, we have compactified $dS^2$ into $S^2$.

The coefficient $Q^2/\gamma$ in front of the $\phi\hat{R}$ term can be regarded 
as an effective inverse topological coupling $1/G_T$. 
It is equal to Newton's coupling $G_N=G_T$. 
This coupling plays an important role in our estimation of the semiclassical entropy 
of two-dimensional de Sitter space.

As is well known, quantum gravity has conformal invariance due to the ambiguity 
in how to separate fields between the background and fluctuations. 
Since scale invariance is a part of the symmetry, 
it may not be very surprising to construct a new solution by a scale transformation. 
To be precise, we have solved the model with an inflaton at the classical level which reproduces 
many features of the solution of exactly renormalized 2D quantum gravity on de Sitter-type space. 
The  introduction of an inflaton is necessary to satisfy the equation of motion 
with respect to the traceless tensor $h^{\mu\nu}$. 
We suspect nature also adopts a similar trick. 
In fact, the classical motion of an inflaton reproduces the quantum effects of $\phi$ in the weak-coupling limit. 

Since the rescaled field $\gamma\phi_c$ obeys the same equation of motion,
the cosmological constant operator keeps the identical expression in the Poincar\'e coordinate: 
\begin{align}
e^{\gamma\phi_c}=\big(\frac{1}{-H\tau}\big)^2.
\end{align}
It is the solution of (\ref{IFTS}) and the classical part of (\ref{ACDU}).
It also satisfies the stationary condition---namely, the coefficient of the linear $\phi$ term vanishes
when the background satisfies $\hat{R}=2H^2$. 
Although there may remain subtle issues in constructing the exact solution 
of two-dimensional de Sitter quantum gravity, the physical picture is robust. 

The most remarkable quantum effect in our solution is that the metric is modified 
and no longer agrees with the volume operator 
\begin{align}
e^{\phi_c}=\big(\frac{1}{-H\tau}\big)^\frac{2}{\gamma}, 
\label{inf}\end{align}
while
\begin{align}
ds^2=dt^2+a^2(t)dx^2,\hspace{1em}a(t)=\big(1+\frac{1-\gamma}{\gamma}Ht\big)^\frac{1}{1-\gamma}.
\label{RNA}\end{align}
The Hubble parameter is
\begin{align}
H(t)=\frac{\dot{a}}{a} = \frac{H}{\gamma}\frac{1}{ 1+\frac{1-\gamma}{\gamma}Ht}.
\label{TDH1}\end{align}

Note that it is no longer constant but it decreases with time. 
The renormalization of the cosmological constant operator with the scaling dimension $\gamma < 1$ 
gives rise to a remarkable result. 
The contribution of matter to the coefficient of the kinetic term of the Liouville field 
reduces the Hubble parameter by a substantial amount. 
Nevertheless, the cosmological constant remains with a definite value. 
The anomalous dimension of the cosmological constant operator has produced a more profound effect. 
The Hubble parameter is no longer constant but decreases with time. 
Let us estimate the acceleration speed of the Universe:
\begin{align}
\dot{H}(t)=\frac{\partial}{\partial t}\big(\frac{\dot{a}}{a}\big) 
=\frac{\ddot{a}}{a}-\big(\frac{\dot{a}}{a}\big)^2
=\frac{\ddot{a}}{a}-H^2(t).
\end{align}
Here, $-\dot{H}(t)$ must be smaller than $H^2(t)$ 
when the expansion of the Universe is accelerating $\ddot{a}>0$. 
From (\ref{TDH1}), these quantities are 
\begin{align}
\dot{H}(t)=-H^2\big(\frac{1-\gamma}{\gamma^2}\big)\big(\frac{1}{1+\frac{1-\gamma}{\gamma}Ht}\big)^2, 
\hspace{1em}H^2(t)=\big(\frac{H}{\gamma}\big)^2\big(\frac{1}{1+\frac{1-\gamma}{\gamma}Ht}\big)^2.
\end{align}
We thus find 
\begin{align}
-\dot{H}(t)=(1-\gamma)H^2(t).
\end{align}
Since $\gamma <1 $ in the semiclassical region where $c>25$, 
the expansion of this class of universes is accelerating. 
The acceleration speed is kept well for weak coupling: 
\begin{align}
-\dot{H}(t)=\frac{1}{Q^2}H^2(t).
\end{align}
Such a universe performs a slow-roll inflation with the following slow-roll parameters:
\begin{align}
\epsilon=\frac{-\dot{H}(t)}{H^2(t)}=\frac{1}{Q^2},\hspace{1em}
\eta=\epsilon-\frac{1}{2}\frac{\ddot{H}(t)}{H(t)\dot{H}(t)}=\frac{2}{Q^2}.
\label{SLPA}\end{align}
On the other hand, the acceleration vanishes at the critical point like
\begin{align}
-\dot{H}(t)=(1-Q)H^2(t).
\end{align}
Note that the classical equation of motion still holds, $\hat{R}=2H^2$. 
This equation appears in the coefficient of the linear term of $\phi\hat{R}$ in (\ref{CQSF}). 
The action for the classical field $\phi_c$ in (\ref{CLLV}) also admits such a solution. 

More precisely speaking, we have separated the classical and quantum parts of this action as follows: 
\begin{align}
\int d^2x  \frac{Q^2}{8\pi}
\big\{-\frac{\gamma}{2}\frac{\partial}{\partial\tau}\phi_c\frac{\partial}{\partial\tau}\phi_c
-\frac{2H^2}{\gamma}e^{\gamma\phi_c}\big\}
\end{align}
\begin{align}
+\int d^2x  \frac{Q_I^2}{8\pi}
\big\{\frac{1}{2}\eta^{\mu\nu}\frac{\partial}{\partial x^\mu}\phi\frac{\partial}{\partial x^\nu}\phi
-2H^2 e^{\gamma\phi_c}(e^{\phi}-\phi-1)\big\}.
\label{QLL}\end{align}
Note that the potential term for the quantum  field $\phi$ is
\begin{align}
V(\phi )=e^{\gamma\phi_c}2H^2\big(\frac{Q_I^2}{8\pi}\big)(e^{\phi}-\phi-1).
\label{STPT}\end{align}
The linear term in $\phi$ vanishes due to the equation of motion for $\phi_c$: 
\begin{align}
\int d^2 x \sqrt{-\hat{g}} \frac{Q_I^2}{8\pi}(\hat{R}-2H^2)\phi\ \rightarrow\ \hat{R} =2H^2. 
\end{align}

We emphasize that the potential has no flat direction as it increases when $\phi\rightarrow\pm\infty$. 
The lifting of the flat direction for negatively large $\phi$ has been achieved 
by demanding that the action be stationary with respect to the de Sitter solution. 
It originates from the $\sqrt{-\hat{g}}\phi \hat{R}$ term in the action. 
Being topological, there is no renormalization of this term 
while the cosmological constant operator $\sqrt{-g}$ is renormalized. 
We have to pay special attention to keep the balance of these two terms 
unless classical solutions are no longer valid. 

We have shown that 
the anomalous dimension $\gamma <1$ of the renormalized cosmological constant operator reduces 
the Hubble parameter $H^2=4\pi\Lambda\gamma/Q^2$ with the fixed cosmological constant $\Lambda$.
This is a short-distance effect of two-dimensional Liouville quantum gravity. 
Furthermore, it makes the Hubble parameter $\dot{a}/a$ time dependent. 
In fact, a negative anomalous dimension implies that it vanishes at a late time. 
The weak-coupling behavior to the leading order of $1/Q^2$ is 
\begin{align}
H(t)\sim {H}\frac{1}{1+\frac{1}{Q^2}Ht}= {H}\frac{1}{1+\frac{1}{Q^2}\log a(t)}.
\label{HBWB}\end{align}
In the perturbation theory, 
time dependence of the Hubble parameter occurs through the IR logarithm $\log a(t)$: 
\begin{align}
H^2(t)\sim H^2\langle e^{\phi}\rangle
\sim H^2\big(1-\frac{2}{Q^2}Ht\big)= H^2\big(1-\frac{2}{Q^2}\log a(t)\big). 
\label{HBWB1}\end{align}
It arises because the momentum integral is logarithmically divergent. 
In the case of exponential expansion of the Universe, 
the one-loop momentum integral behaves as $\log a(t)$ 
as the infrared cutoff goes like $L/a(t)$ for a fixed UV cutoff $L$. 
The negative sign of the one-loop correction is due to the negative sign of the kinetic term of $\phi$. 
The shielding effect of the cosmological constant does not occur if the metric is positive. 
The exact solution is in accord with the perturbation theory 
in important issues whose significance is still to be explored. 

The Hubble parameter $H(t)$ decays inversely proportional to $\log a(t)$---namely, 
the e-folding number at a late time. 
It decays faster when the effective coupling $1/Q^2$ is stronger. 
We show that there are no other pure IR effects which diffuse $H(t)$. 
This important conclusion is obtained from the investigation of the exactly renormalized Lagrangian. 
Since it is a conclusive result on the fate of the Hubble parameter $H(t)$ 
in two-dimensional Liouville quantum gravity, 
we briefly recall its renormalization procedure in Appendix B.

We have suspected that the inflaton may be a dual description of quantum effects in gravity. 
It is encouraging that this two-dimensional toy model  provides us a concrete example of such an idea. 
Let us go back to the original action before eliminating the inflaton by the equation of motion: 
\begin{align}
\int d^2x \frac{1}{8\pi}\sqrt{-\hat{g}}	
\big\{\frac{Q^2}{2}\hat{g}^{\mu\nu}\partial_\mu\phi\partial_\nu\phi +2\hat{R}\phi
-\frac{1}{2} \hat{g}^{\mu\nu}\partial_\mu f \partial_\nu f -2H^2Q^2e^{\phi-(1-\gamma)f}\big\}, 
\label{IFAT2}\end{align}
where we took the weak-coupling limit. 
The effective Hubble parameter is induced by putting the classical solution of the inflaton into the potential: 
\begin{align}
H^2(f)=H^2e^{-\frac{1}{Q^2}f}=H^2e^{-\frac{1}{Q^2}2Ht}. 
\end{align}
The slow-roll parameters agree with the estimate of the exact solution (\ref{SLPA}): 
\begin{align}
\epsilon&= Q^2\big(\frac{V'}{V}\big)^2=\frac{-\dot{H}(t)}{H^2(t)}=\frac{1}{Q^2}, \notag\\
\eta&=2Q^2\frac{V''}{V}=2\epsilon+2Q^2(\frac{V'}{V})'=\frac{2}{Q^2}.
\end{align}
At weak coupling, our solution is a slow-roll inflation model 
where the inflaton $f$ rolls down the exponential potential
\begin{align}
2H^2Q^2\exp\big(-\frac{1}{Q^2}f\big).
\end{align}
We find it remarkable that this toy model underwrites a long-suspected scenario 
that an inflaton is to provide a dual description of quantum effects in gravity. 

There exist a class of slow-roll inflation models and Liouville gravity 
which are connected by a rotation of the conformal mode and the inflaton \cite{MM}. 
Our proposal is a duality between Liouville gravity and semiclassical inflation models. 
In fact, we point out that these formally identical models are dual to a unique inflation model in Appendix C.
We have introduced an inflaton to solve the equation of motion 
with respect to the traceless mode of the metric $h^{\mu\nu}$ in Liouville gravity. 
Remarkably, the classical behavior of the inflaton reproduces known quantum effects in Liouville gravity 
in the weak-coupling region. 
In view of the proliferation of inflation models, it is  a very attractive possibility 
that a unique inflaton model appears out of quantum effects of Liouville gravity or even Einstein gravity. 

We regard (\ref{IFAT2}) as a low-energy effective theory just like pions in QCD. 
In other words, we investigate this theory at tree level to avoid double counting of quantum effects. 
As an example, we may examine the density perturbation in this model following the standard prescription. 
The inflaton field may fluctuate around the classical solution as 
\begin{align}
f_c(t)+f = f_c(t+\delta t) .
\end{align}
We pick a comoving gauge to eliminate the fluctuation of the inflaton, 
\begin{align}
\delta t =\frac{f}{\dot{f}_c(t)}. 
\end{align}
It then generates density perturbation, 
\begin{align}
-dt^2+e^{2Ht}e^{2\zeta}dx^2,\hspace{1em}\zeta=-H\delta t=-H\frac{f}{\dot{f}_c(t)}. 
\end{align} 
The spectrum of the density perturbation is
\begin{align}
\langle\zeta_{\vec{k}}\zeta_{\vec{k}'}\rangle=(4\pi)^2 \delta (\vec{k}+\vec{k}') 
\frac{1}{2k}\big(\frac{H}{\dot{f_c}}\big)^2. 
\end{align}
In our case $\dot{f_c}=H$, so there seems to be no enhancement: 
\begin{align}
\langle\zeta_{\vec{k}}\zeta_{\vec{k}'}\rangle=(4\pi)^2\delta (\vec{k}+\vec{k}') \frac{1}{2k}.
\end{align}
However, it is enhanced in comparison to the conformal mode: 
\begin{align}
\langle\phi_{\vec{k}}\phi_{\vec{k}'}\rangle=-(4\pi)^2 \delta (\vec{k}+\vec{k}') \frac{1}{2k}\frac{1}{Q^2}.
\end{align}
So the enhancement of the density perturbation over the gravitational modes 
by a slow-roll parameter $\epsilon$ appears to hold also in two-dimensional de Sitter space. 

The conclusion in this section is that the renormalized volume operator $e^{\gamma\phi}$ is obtained 
after integrating short-distance degrees of freedom. 
It is the relevant operator to investigate long-distance physics. 
The scaling dimension $\gamma$ is less than canonical $\gamma<1$ in the semiclassical region where $c> 25$.
We have examined a de Sitter-type solution of the renormalized Liouville action. 
It shows that the Hubble parameter becomes not only time dependent but vanishes at a late time. 
This effect clearly breaks de Sitter invariance 
and is caused by the renormalization of the cosmological constant operator. 

\section{Entropy production at the horizon diffuses cosmological constant}
\setcounter{equation}{0}

We recall that the semiclassical entropy in four-dimensional de Sitter space is given by
\begin{align}
S_0=\frac{A_0}{4G_N}=\frac{\pi}{H^2G_N}. 
\end{align}
It can be compared with our estimate (\ref{SCEP}) in two-dimensional de Sitter space:
\begin{align}
S_0=\frac{1}{G_N}\log \frac{H_0^2}{H^2(t)}. 
\end{align}

The semiclassical entropy of the system is given by $Q^2/\gamma \sim c/6$, 
which plays the role of the inverse Newton's coupling $1/G_N$. 
So the increase of entropy by adding more matter reduces the Hubble parameter. 
Entropy also increases if the Hubble parameter decreases. 
The trend is in the same direction with four-dimensional de Sitter space 
although the speed of the increase is much slower:
logarithmic $\log(1/H^2(t))$ versus power law $1/H^2(t)$. 
In this section, we investigate IR effects on the Hubble parameter from an entropic point of view. 

Due to the existence of the cosmological horizon, plain waves constantly accumulate at it. 
They are called superhorizon modes and are constant in space. 
They do change with time over the cosmic scale. 
For a static observer inside, more and more constant modes are accumulating at the horizon. 
As they evolve in a stochastic process, it is expected that entropy is continuously generated there. 
Simultaneously, the Hubble parameter is diffused with the evolution of the Universe. 
We show that such a dramatic process takes place in two-dimensional Liouville quantum gravity.
This conclusion follows from the entropy generation effects 
associated with the evolution of superhorizon modes. 

The precursor of the effect is the quantum fluctuation of the conformal mode
\begin{align}
\langle e^{\phi} \rangle = \langle e^{\phi_c+\gamma\tilde{\phi}} \rangle \sim e^{\phi_c(t)+\frac{1}{2}\gamma^2\langle \tilde{\phi}^2 \rangle}. 
\end{align}
Here $\phi_c(t)$ denotes a classical solution while
\begin{align}
\langle 
\tilde{\phi}^2 \rangle=-\frac{4}{Q^2}\int_{P_\text{min}}^{P_\text{max}} \frac{dP}{P}=-\frac{4}{Q^2}\log a(t).
\end{align}
In this integral with respect to physical momentum, we fix the UV cutoff $P_\text{max}\sim L$. 
We identify the IR cutoff as $P_\text{min}=L/a(t)$. 
Here $a(t)=\exp(\phi_c(t)/2)$ is the scale factor of the Universe and $1/L$ is the initial size of the Universe. 
Since we consider the conformal zero mode $\phi_0$, it can only depend on time. 
Its characteristic timescale is the Hubble scale. 

In this way, the quantum IR fluctuation grows: 
\begin{align}
\langle \tilde{\phi}^2 \rangle\sim-\frac{2}{Q^2}\phi_c (t)\ \Rightarrow\ 
\langle e^\phi\rangle \sim e^{(1-\frac{\gamma^2}{Q^2})\phi_c (t)}.  
\label{LCC}\end{align}
This effect may diminish the effective cosmological constant as the Universe expands \cite{KK}: 
\begin{align}
H^2_\text{eff} \sim H^2e^{-\frac{\gamma^2}{Q^2}\phi_c (t)}\sim H^2a(t)^{-\frac{2\gamma^2}{Q^2}}. 
\label{EFCC}\end{align}
The important point here is that the quantum IR effect is time dependent 
and hence cannot be subtracted by a dS-invariant counterterm. 
We introduce counterterms in accordance with the general coordinate invariance of the action. 
If the background de Sitter space is stable, the general coordinate invariance reduces to the dS invariance. 
On the contrary, 
a nontrivial anomalous dimension of the cosmological constant operator spoils the dS invariance. 
The scale-invariant de Sitter solution in (\ref{URSL})--(\ref{metric}) is replaced 
by the inflation-type solution in (\ref{inf})--(\ref{RNA}). 
Furthermore, an inflaton field has emerged out of the necessity to satisfy the equation of motion (\ref{HEQ}). 

In (\ref{EFCC}), we have only considered the leading-order IR effect in $H^2$. 
Note that this one-loop IR effect (\ref{EFCC}) is consistent to the leading order with the prediction (\ref{TDH1}) 
based on the  exact scaling dimension $\gamma$ of the cosmological constant operator: 
\begin{align}
H^2(t)= \frac{H^2}{\gamma}(1+\frac{1-\gamma}{\gamma}Ht)^{-2}\sim H^2(1-2\frac{1-\gamma}{\gamma}Ht)
\sim H^2(1-\frac{1}{Q^2}2Ht). 
\label{TDH2}\end{align}
So, it would be a double counting to take into account 
both the scaling due to the anomalous dimension and the IR logarithm. 

Let us examine what creates the entropy to reduce the Hubble parameter as above.
We conjecture that de Sitter entropy is carried by a conformal zero mode. 
It performs a Brownian motion due to the constant disturbance by newcomers that have just joined it. 
Such a process can be investigated by a Fokker-Planck-type equation 
which governs the evolution of the distribution function of the conformal zero mode  $\rho(\phi_0)$. 
To be more precise, de Sitter entropy is the von Neumann entropy of $\rho(\phi_0)$. 

We may put this formula (\ref{TDH2}) into the semiclassical estimate of the de Sitter entropy: 
\begin{align}
\frac{Q^2}{\gamma}\log\frac{1}{H^2(t)}
&=\frac{Q^2}{\gamma}2\log(1+\frac{1-\gamma}{\gamma}Ht) \notag\\
&\sim\frac{Q^2}{\gamma}\frac{1-\gamma}{\gamma}2Ht = 2Ht. 
\label{SCEN}\end{align}
The speed of entropy generation is given by taking the time derivative of the above:
\begin{align}
\frac{2H}{(1+\frac{1-\gamma}{\gamma}Ht)} = 2\gamma H(t) .
\label{SCEN1}\end{align}
It is given by the Hubble parameter and thus it also slows down with cosmic evolution. 
This semiclassical estimate can be compared 
with that of the von Neumann entropy of the distribution function $\rho(\phi_0)$. 

The distribution functions of Fokker-Planck equations are well approximated by the Gaussian for weak coupling.
(\ref{ADEN}) shows that there is a $-(1/2) \log \omega$ term in the von Neumann entropy. 
$1/\omega$ is the standard deviation of the Gaussian, and the entropy grows as $\omega$ decreases. 
(\ref{SCEN1}) implies 
\begin{align}
-\frac{1}{2}\frac{\partial}{\partial t}\log\omega \sim 2H, 
\label{SCEN2}\end{align}
which is in qualitative agreement with (\ref{EVIB}), 
which is the increasing speed of the von Neumann entropy of $\rho(\phi_0)$ under the Fokker-Planck equation. 

We believe that carrying out the renormalization and summing IR logarithms by Fokker-Planck equations
is double counting; we should not do them both.
We should only perform the renormalization that is necessary anyway. 
We thus conclude that (\ref{SCEN}) is the correct semiclassical estimate. 
The new entropy is generated by the accumulation of conformal zero modes. 
They manifest as IR logarithms in perturbation theory which grows with time. 
Fortunately, in two dimensions, we can decipher their physical effects through renormalization procedures 
since UV and IR effects are closely related. 

Geometric entropy of de Sitter space arises since there is a much wider world outside the cosmological horizon. 
It is analogous to the entangled entropy 
in the sense that both arise after integrating out the Hilbert space of the outer world. 
From the viewpoint of observers inside the cosmological horizon, 
they see nothing going out of the horizon. 
For them, only conformal zero modes are piling up. 
So they must carry the entire de Sitter entropy and this investigation supports such a point of view. 
We are able to verify that conformal zero modes contribute to shield the Hubble parameter 
due to its negative sign for the kinetic energy. 
Simultaneously, we can offer various evidences that they generate de Sitter entropy 
at a rate in accord with semiclassical estimates. 

We recall here
\begin{align}
\frac{\gamma^2}{4}\langle\phi^2\rangle=-\gamma^2\frac{1}{Q^2} \log a
\sim -\frac{1}{Q^2}\gamma Ht =-\frac{1-\gamma}{\gamma}Ht.
\end{align}
This is because of the relations in (\ref{saddle}),  
\begin{align}
\frac{1-\gamma}{\gamma}=\frac{\gamma}{Q^2}, 
\end{align}
and (\ref{RNA}), 
\begin{align}
\log a (t)=\frac{1}{1-\gamma}\log (1+\frac{1-\gamma}{\gamma}Ht). 
\end{align}
The expectation value of any function of $\gamma\phi$ must be a function 
of $\frac{1-\gamma}{\gamma}Ht$ as the following relation holds: 
\begin{align}
\langle(\gamma\phi)^2\rangle=-4\log(1+\frac{1-\gamma}{\gamma}Ht). 
\end{align}
The time dependence of the Hubble parameter $H(t)$ implies that 
the lower cutoff of the momentum integral is the inverse of the size of the Universe for the fixed UV cutoff. 
It clearly originates from the IR effects. 
Since this factor $\frac{1-\gamma}{\gamma}Ht$ is the leading log, we need to sum all powers of this variable. 
The solutions of the renormalized action are such functions. 
In this respect, we believe that the leading IR logs are already contained in them. 

The anomalous dimensions are the short-distance effect. 
However, it also predicts a long-distance cutoff dependence of the operator 
since the short-distance and long-distance cutoffs must appear together as the ratio on dimensional grounds. 
This is because the propagators of the minimally coupled scalars are the same 
in both the UV and IR regions in two dimensions.
We have thus confirmed that the time dependence of the cosmological constant operator is related 
to the anomalous dimension $:e^{\phi}:=e^{\gamma\phi_c}$ to the leading order in $1/Q^2$, 
where $:e^{\phi}:$ denotes the renormalization. 
It is reasonable to believe that they also contain all leading log effects. 
The exact expression shows that the anomalous dimension $\gamma-1$ is negative 
in the semiclassical regime $c > 25$. 
Surprisingly, the short-distance effect alone makes the Hubble parameter time dependent. 
(\ref{TDH1}) further shows that two-dimensional de Sitter space is doomed 
as the Hubble parameter $H(t)$ fades away with $\gamma <1$. 
In the weak-coupling limit, $H(t)$ decays as $Q^2/t$ with cosmic time. 
Nevertheless, it is important to investigate if there are other sources of IR logarithms 
in the entire Liouville theory.  

The loop integral is logarithmically divergent with respect to the IR cutoff. 
It is also known that the $n$ th powers of IR logarithms may appear 
if the diagram contains $n$ propagators \cite{Weinberg}. 
Leu us recall that each logarithm behaves as
\begin{align}
\frac{1}{Q^2}\log a(t)\sim \frac{1}{Q^2} Ht. 
\end{align}
So it becomes $O(1)$ if the e-folding number of the Universe becomes $O(Q^2)$. 
We thus need to sum up all of them at late times. 
The leading IR logarithms of these origins can be summed up by the Langevin and Fokker-Planck equations. 
In two dimensions, the effective gravitational coupling is  $1/Q^2$. 
It is very large in comparison to four dimensions 
where $1/Q^2$ is replaced by the notorious ratio $(H/M_P)^2\sim 10^{-120}$, where $M_P$ is the Planck mass. 
Nevertheless, such effects may have a significant impact on the evolution of the Universe. 
Fortunately, this problem turns out to be solvable by renormalizing the cosmological constant operator exactly. 

In order to understand the geometric entropy of the two-dimensional de Sitter space 
from superhorizon degrees of freedom, 
it is useful to investigate them in Liouville gravity. 
Let us recall the definition of entropy: 
\begin{align}
\beta F = -\log Z,\hspace{1em}S=-(1-\beta\frac{\partial}{\partial \beta})\beta F, 
\end{align}
where $\beta=1/T$. 
The two-dimensional de Sitter space may be rotated into $S^2$ 
and $\beta$ corresponds to the radius $l$ of $S^2$ as $\beta=2\pi l$. 
If we assume scale invariance, $\beta F$ cannot depend on $\beta$ since it is dimensionless. 
Therefore, the conformal anomaly and Liouville action are the source of the nontrivial geometric entropy. 
Since the size of the Universe is dynamically determined in quantum gravity, 
$-\beta F=\log Z$ gives us nothing but entropy. 
It is stationary with respect to a change of the geometry of the manifold such as $\beta$. 

Note that (\ref{SCEP}) is reminiscent of the entangled entropy 
with the central charge $c$ \cite{Wilczek,Takayanagi}: 
\begin{align}
S_\text{en}=\frac{c}{3}\log\frac{b}{a}, 
\label{Entm}\end{align}
where $a$ and $b$ denote the short-distance and long-distance cutoffs of the subsystem respectively. 
We may identify $H_0/H(t) =b/a$. 
In conformal field theory, 
we cannot associate any dimensionfull parameters with $H_0$ or $b$ since there are none. 

The geometric entropy may be the quantized version of the entangled entropy. 
Entangled entropy is obtained from the density matrix of the subsystem. 
It is the entropy of the mixed state after integrating local degrees of freedom belonging to the outer system. 
The geometric entropy of de Sitter space is expected to be constructed in an analogous way. 
The density matrix may be obtained by integrating out the states outside the horizon. 
The expectation value of the operators inside the horizon can be evaluated by the density matrix. 
In field theory, this may be accomplished by evaluating correlation functions in the Liouville gravity. 
In the case of conformal zero modes, their correlators are calculable 
from the Fokker-Planck-type distribution function $\rho(\phi_0)$. 
Understanding the relations of these various approaches will shed light on elucidating this problem. 

Let us recall how to estimate the entangled entropy. 
We may divide the real line into two sectors: positive and negative half-lines. 
We may change coordinates from the plane to a cylinder $z=e^w$. 
The lower half-plane is mapped to a rectangular region 
where the the lower line segment corresponds to our section 
and the upper line section corresponds to the outer section. 
We may impose periodic boundary conditions on the remaining sections. 
The density matrix is obtained by integrating out the fields on the outer segment 
after we glue two cylinders together.
 
In this case, the problem is effectively compactified onto a torus 
while geometric entropy of $dS^2$ is often studied by compactifying it onto $S^2$.
After integrating out the localized states outside the cosmological horizon, 
we are left with a half-line of the length $2l$. 
It becomes a circle if we adopt periodic boundary condition on this strip. 
It is natural and may be even the right choice to compactify $dS^2$ to $S^2$ 
with the identification of this circle and the circumference. 
The density matrix $\rho (\phi,\phi')$ may be obtained by performing the path integral of the fields 
like the conformal mode on $S^2$ with a specified field $\phi,\phi'$ at both sides of the equator. 
The expectation value of the fields may be evaluated by inserting them on the equator 
and performing the path integral on the whole $S^2$ with a suitable action like Liouville quantum gravity. 
The geometric entropy $S$ can be evaluated by simply evaluating the partition function $Z$ 
since it gives the geometric entropy $S=\log Z$ right away in quantum gravity. 
Suppose the Hubble parameter changes slowly with cosmic time. 
In this case, it may be a good strategy to change the radius of the corresponding $dS^2$ 
and $S^2$ as $1/H(t)$. 
As far as conformal zero mode is concerned, there is no problem in Euclidean rotation 
since the potential term dominates the kinetic term. 

So far we have assumed that the matter system is at the critical point---i.e., conformally invariant. 
In a more generic situation, the central charge $c$ is known to be a decreasing function 
with respect to the IR cutoff and hence time $\phi_c (t)$. 
For example, a nonlinear sigma model may develop a mass gap. 
In such a situation, the number of massless scalar fields decreases. 
This effect may enhance the magnitude of the anomalous dimension 
and the screening effect of the cosmological constant. 

The conformal zero mode performs a Brownian motion with the scale set by $1/Q^2$.
As the Universe expands, plane waves constantly come out of the horizon to join the superhorizon mode. 
They collide with the main body just like a Brownian process of the strength $1/Q^2$. 
It is noteworthy that the metric of the conformal mode is negative like Einstein theory in four dimensions, 
although there could be an equilibrium distribution for the conformal zero mode 
if there is a countereffect to diffusion. 
However, there is no such possibility here since we have no drift force 
due to the uniqueness of the classical solution. 
See Appendix D for its explanation. 

We focus on the dynamics of the superhorizon mode of the conformal factor of the metric. 
Its cosmic evolution in real spacetime can be investigated by a Langevin-type equation. 
The ensemble average of a function of $f(\phi(t))$ is a natural observable in the system 
governed by a Langevin equation. 
The Langevin equation is equivalent to the Fokker-Planck equation. 
We define the ensemble average of a function of $f(\phi(t) )$ as
\begin{align}
\langle f(\phi (t))\rangle=\lim_{n\rightarrow\infty}\frac{1}{n}\sum_{i=1}^{n} f(\phi_i(t)), 
\end{align}
where $i$ denotes the observation of the $i$ th member. 
In this context, it is natural to introduce a distribution function $\rho_t(\phi )$ in the following way: 
\begin{align}
\langle f(\phi(t) )\rangle=\int d\phi \rho_t (\phi )f(\phi ). 
\end{align}
The probability distribution function $\rho_t (\phi)$ obeys the Fokker-Planck equation. 
This formalism is close to the field theory approach 
especially with respect to investigating the superhorizon mode.
The system may approach an equilibrium state at a late time. 
In that case, $\rho(\phi)$ describes an equilibrium state whose temperature $T$ is determined 
by the strength of the random force. 
We thus conclude that
\begin{align}
\rho(\phi) = e^{-\beta V (\phi))}\frac{1}{Z}, 
\end{align}
where $\log Z = -\beta F$.
From this formula, we can verify that 
the von Neumann entropy of $\rho$ gives us the entropy of this equilibrium state:  
\begin{align}
\int d \phi \left(- \rho\log \rho \right)= 
\int d \phi  (\beta E -\beta F)\rho =S.
\end{align}

Although our strategy is to investigate IR effects in real spacetime with Langevin equation, 
Fokker-Plank equations relate the problem with thermodynamics. 
The equilibrium state is studied very well  by Euclidean field theory. 
We can estimate the von Neumann entropy of $\rho_t$ even if it is not equilibrated. 
It is the measure of the entropy of the system 
which evolves according to the Fokker-Planck or Langevin equation. 

In order to connect the shielding mechanism of the cosmological constant 
with entropy generation at the horizon, 
we employ a stochastic approach \cite{Starobinsky,Woodard2005}. 
The Langevin equation for the conformal zero mode  $\phi_0$ with respect to the cosmic time $t$ is derived 
in Appendix D. 
The superhorizon mode of the conformal degree of metric is given by 
\begin{align}
\phi_0(x)=\sqrt{\frac{8\pi}{Q^2}}\int \frac{d\vec{p}}{2\pi}\theta (Ha(t)-p)
\big(a_{\vec{p}}\frac{1}{\sqrt{2p}}e^{i\vec{p}\cdot\vec{x}}
+a_{\vec{p}}^\dagger\frac{1}{\sqrt{2p}}e^{-i\vec{p}\cdot\vec{x}}\big), 
\end{align}
where $[a_{\vec{p}},a_{\vec{p}'}^\dagger]=-2\pi\delta(\vec{p}-\vec{p}')$. 
Since plane waves become constant in time, 
the time dependence is caused by the step function which restricts physical momenta $P<H$. 
The Langevin equation is given by 
\begin{align}
\dot{\phi}(x) = \dot{\phi}_0 (x),\hspace{1em}
\langle\dot{\phi}_0 (t,\vec{x})\dot{\phi}_0 (t',\vec{x})\rangle=-\frac{4}{Q^2}H\delta (t-t'). 
\end{align}
We have dropped the drift term but kept the quantum fluctuation effect. 
Our purpose in this investigation is not to do double counting 
as the renormalization of the cosmological constant operator occurs by identical quantum fluctuations. 
It is rather to see to what extent we can reproduce the features of the exact solutions. 
By doing so, we shall be able to examine the consistency of our understanding on this issue. 

If $\phi (t)$ obeys the Langevin equation, 
the Fokker-Planck equation for the distribution function $\rho_t(\phi )$ follows 
\begin{align}
\dot{\rho} = -\frac{2}{Q^2}H\frac{\partial}{\partial \phi^2}\rho. 
\end{align}
We notice that the right hand-side is of the opposite sign in comparison to those appearing in
the study of unitary matter systems. 
Of course, this is due to the negative metric of the conformal mode. 
So this equation is obtained by the time reversal of the former.
It appears that our equation listed so far in this section runs the show backward
in comparison to the standard evolution in the matter system.
 
However, there is an important issue we have to address in quantum gravity. 
The distribution of conformal zero modes $\phi$ must change under the evolution. 
For this reason, we may include the renormalization factor $\omega$ 
for the cosmological constant operator. 
Note that the linear term in $\phi$ cancels in the potential 
which ensures that the equation of motion $\hat{R}=2H^2$ holds: 
\begin{align}
V= \sqrt{-\hat{g}}H^2\frac{Q^2}{\omega}(e^{\omega\phi}-\omega\phi-1). 
\end{align}
We assume under Euclidean rotation 
\begin{align}
\frac{i}{4\pi}\int d^2x \sqrt{-\hat{g}}H^2\frac{Q^2}{\omega}(e^{\omega\phi}-\omega\phi-1)\ \rightarrow\ 
\frac{Q^2}{\omega}(e^{\omega\phi}-\omega\phi-1), 
\label{ECPT}\end{align}
when we compactify $dS^2$ into $S^2$ with the radius of $1/H$. 
As we emphasized in the preceding section, 
the renormalization properties of the operators $\sqrt{-g}$ and $\sqrt{-\hat{g}} \phi \hat{R}$ are different 
while the de Sitter background is realized by balancing them. 
In fact, the $e^{\omega\phi}$ and $\phi$ terms in the potential (\ref{ECPT}) come 
from the former and latter operators respectively.\footnote{
The identity is our normalization convention.} 
It is required to keep the balance of the two different operators.
We need a formalism which lets the renormalization of operators cancel 
the effect of the evolution by the Fokker-Planck equation. 

We thus assume the following distribution containing ${\omega}$: 
\begin{align}  
\rho_{\omega}=N_\omega e^{-\frac{Q^2}{\omega}(e^{\omega\phi}-\omega\phi-1)}
\sim \sqrt{\frac{Q^2\omega }{2\pi}}e^{-\omega Q^2\frac{1}{2}\phi^2}, 
\label{SLMF}\end{align}
where $N_\omega$ is the normalization factor. 

The readjustment of the background under time evolution is realized by requiring
time independence for the distribution function: 
\begin{align}
\dot{\rho}_\omega=-2H\frac{1}{Q^2}\frac{\partial ^2}{\partial \phi^2}  \rho_\omega
+\dot{\omega}\frac{\partial}{\partial \omega}\rho_\omega
=0.
\end{align} 
In this way, the background adjusts itself automatically 
to cancel the evolution brought by the Fokker-Planck equation. 

The gravitational Fokker-Planck equation is 
\begin{align}
\dot{\omega}\frac{\partial}{\partial \omega}\rho_{\omega}
=2H \frac{1}{Q^2}\frac{\partial ^2}{\partial \phi^2} \rho_{\omega}. 
\label{MDFP}\end{align}
After the dust settles, the sign of the right-hand side turns back to normal. 
The time derivative of the distribution function is specified through its $\omega$ dependence. 

From this modified Fokker-Planck equation (\ref{MDFP}), 
we obtain the following evolution equation for the background $\omega$: 
\begin{align}
\dot{\omega}=-4H\omega^2. 
\end{align}
The solution is 
\begin{align}
\omega(t)=\frac{1}{1+4Ht}. 
\label{FPSL}\end{align}

The initial probability distribution $\rho_0$ is 
\begin{align}
\rho_0=N_0 \exp \big\{-Q^2(e^{\phi}-\phi-1)\big\}
=N_0 \exp\big\{-\frac{4\pi}{H^2}\Lambda (e^{\phi}-\phi-1)\big\}.
\label{FDM}\end{align}
This formula suggests a Euclidean system on $S^2$ with a radius $1/H$. 
This solution coincides with our original potential before the effects of IR logarithms 
become important---namely, at the beginning of the de Sitter expansion. 
In the semiclassical region where $Q^2$ is large, this potential is well approximated by a Gaussian: 
\begin{align}
N_0\int d\phi \exp\big\{-Q^2(e^{\phi}-\phi-1)\big\}
\sim \frac{Q}{\sqrt {2\pi }}\int d\phi \exp ( - \frac{Q^2}{2}\phi^2). 
\end{align} 

We have shown that there is an effect to reduce the effective cosmological constant (\ref{EFCC}). 
As long as this effect is concerned, we believe that the UV investigation (\ref{TDH1}) has shown that 
the Hubble parameter acquires time dependence and it eventually vanishes. 
Since the propagators are identical in both the UV and IR regions, 
the two birds can be dealt with by a single stone. 
We argue there are no drift force effects since the solution of the theory is unique. 
There are unstable deformations if they increase the entropy of the system. 
As we emphasized, the system evolves toward the state with maximum entropy in quantum gravity. 

What is the geometric entropy of de Sitter space? 
We propose that it is the entropy of the superhorizon conformal mode 
which accumulates with cosmic expansion. 
There are no other massless modes in two-dimensional Liouville gravity. 
Furthermore, the entropy could increase in a stochastic process. 
Let us evaluate the von Neumann entropy of the conformal zero mode with (\ref{FDM}): 
\begin{align}
S_0&=-tr \rho_0\log\rho_0 \notag \\
&= \int d\phi  \rho_0\big\{{Q^2}(e^\phi-\phi-1)-\log Q+\frac{1}{2}\log(2\pi)\big\} \notag\\
&\sim \frac{1}{2} -\log Q+\frac{1}{2}\log(2\pi) .
\end{align}
In our view, the von Neumann entropy of the superhorizon mode is 
the identity of the geometric entropy of de Sitter space. 
A characteristic feature of this expression is its $Q^2$ dependence. 
The von Neumann entropy becomes larger 
if the effective gravitational coupling $1/Q^2$ becomes stronger.\footnote{
We do not exclude the possibility that a constant term like $Q^2$ is missing 
since it becomes negative for large $Q$.}

Let us investigate its $Q^2$ dependence from the two-dimensional Liouville quantum gravity point of view: 
\begin{align}
\frac{\partial}{\partial Q^2} \log Z = -\langle(e^{\phi}-\phi-1)\rangle=-\frac{1}{2Q^2}. 
\label{Q2DP}\end{align}
Here $Z$ is the partition function of the superhorizon sector of two-dimensional Liouville gravity on $S^2$: 
\begin{align}
Z=\int d\phi e^{-Q^2(e^{\phi}-\phi-1)}. 
\end{align}
Note that the potential is bounded below 
and the expectation value of the $n$-point function of the superhorizon mode is calculable. 
The potential dominates the kinetic energy for the superhorizon conformal mode. 
We can safely ignore the wrong-signed kinetic term in comparison to the potential term. 
The wrong sign problem of the conformal mode may turn out be a blessing 
with respect to the cosmological constant problem. 
As is explained, $\log Z$ gives us the entropy itself in quantum gravity. 
So a $Q^2$ dependence of von Neumann entropy is consistent 
with a geometric entropy of the superhorizon conformal modes of Liouville gravity (\ref{Q2DP}). 
The expectation value of any function $f(\phi )$ is well defined 
unless $f(\phi )$ grows too rapidly as $\phi\rightarrow\pm\infty$ to spoil the convergence of the integral. 
The expectation value of $f(\phi )$ is real if $f$ is a real function. 

We consider the change of von Neumann entropy in the stochastic process
by introducing one parameter deformation of an initial distribution function by ${\omega}$.
This factor allows us to renormalize the cosmological constant operator as 
\begin{align}
\rho_\omega =N_{\omega} e^{-\frac{Q^2}{\omega}(e^{\omega\phi}-\omega\phi-1)}. 
\label{ADEF}\end{align}
In a Gaussian approximation,
\begin{align}
N_{\omega}=\frac{Q\sqrt{\omega}}{\sqrt{2\pi}}. 
\end{align}
We rotate the Minkowski space-time potential of $dS^2$ into Euclidean $S^2$: 
\begin{align}
\frac{i}{4\pi} \int d^2x \sqrt{-\hat{g}}H^2\frac{Q^2}{\omega} (e^{\omega\phi}-\omega\phi-1)\ \rightarrow\ 
\frac{Q^2}{\omega} (e^{\omega\phi}-\omega\phi-1). 
\end{align}
The corresponding von Neumann entropy is 
\begin{align}
S_\omega&=-tr \rho_\omega\log\rho_\omega \notag\\
&=\int d\phi \rho_\omega\big\{\frac{Q^2}{\omega}(e^{\omega\phi}-\omega\phi-1)
-\log Q-\frac{1}{2} \log \omega+\frac{1}{2}\log(2\pi)\big\} \notag\\
&\sim S_0-\frac{1}{2}{\log \omega}. 
\label{ADEN}\end{align}

This expression shows that entropy increases if $\omega$ decreases from the initial point $\omega=1$. 
So there might be an unstable deformation of this configuration. 
If $\omega$ decreases, the distribution of the superhorizon mode spreads out. 
Let us examine a possible instability of this configuration in the vicinity of $\omega\sim 1$ 
under the Fokker-Planck equation: 
\begin{align}
\dot{\omega}\frac{\partial}{\partial \omega}S_\omega
&=-tr \dot{\omega}\frac{\partial}{\partial \omega}\rho_\omega \log \rho_\omega \notag\\
&=-tr\frac{2}{Q^2}H\frac{\partial^2}{\partial \phi^2}\rho_\omega \log \rho_\omega \notag\\
&\sim 2H\omega
=-\frac{1}{2\omega}\dot{\omega}. 
\label{EVIB}\end{align}
From the inspection of (\ref{EVIB}), it is clear that 
the von Neumann entropy of $\rho_\omega$ always increases. 
In particular, its growth $\Delta S=2Ht$ 
as the system evolves away from the initial distribution with $\omega=1$. 

This result is consistent with semiclassical estimates of the geometric entropy (\ref{SCEN}) 
when $\omega\sim 1$.
The exact solution in the weak-coupling region shows that the entropy increases as
\begin{align}
-\frac{1}{2}\log \omega =2\frac{Q^2}{\gamma}\log\big(1+\frac{1-\gamma}{\gamma}Ht\big)\sim 2Ht 
\end{align}
for the weak-coupling or short-time limit. 
We can verify that von Neumann entropy increases logarithmically 
in the evolution under the Fokker-Planck equation by using the explicit solution (\ref{FPSL}): 
\begin{align}
S_\omega=\frac{1}{2}\log (1+4Ht)\sim 2H t. 
\end{align}
All approaches agree that entropy grows as $2Ht$ away from the initial distribution function. 
They also agree that $H(t)$ eventually vanishes. 
The eventual fate of 2D de Sitter space is not agreed upon. 
The exact solution predicts it is $Q^2$ dependent. 
The slow-roll parameter is given by $\epsilon=1/Q^2=\eta/2$. 
It also predicts that the acceleration stops at the critical point $Q^2=0$. 
The Fokker-Planck equation predicts a more rapid slowdown of the acceleration. 

The existence of configurations of higher von Neumann entropy implies that 
the potential for the superhorizon modes of two-dimensional Liouville gravity is modified 
also as in a process of evolution: 
\begin{align}
V(\phi)=\frac{Q^2}{\omega}(e^{\omega\phi}-\omega\phi-1). 
\end{align}
The partition function for the superhorizon sector of conformal mode evolves as
\begin{align}
Z_\omega=\int d\phi e^{-\frac{Q^2}{\omega}(e^{\omega\phi}-\omega\phi-1)}. 
\end{align}
This is because the $\phi$ field obeys the identical Langevin equation in two-dimensional gravity. 
So the potential for the conformal mode must change according to the Fokker-Planck equation; 
it must be identical to that of the distribution function (\ref{ADEF}). 
With this potential, we can reproduce the $\omega$ dependence of von Neumann entropy from Liouville gravity: 
\begin{align}
\log Z_\omega \sim -\log Q-\frac{1}{2}\log \omega. 
\end{align}
So geometric entropy in Liouville gravity in de Sitter space is consistent 
with the von Neumann entropy (\ref{ADEN}) of superhorizon modes. 
From these considerations, we are able to obtain consistent pictures of these IR effects 
on the Hubble parameter. 
The Hubble parameter is generically suppressed by the e-folding number. 
The inhabitants of two-dimensional de Sitter space may always wonder 
why the Hubble scale is always the size of the Universe. 

Let us check universes like those given by (\ref{FPSL}). 
$\omega$ is just like the scaling dimension of the cosmological constant $\gamma$. 
This Universe starts with a slow-roll inflation while the slow-roll parameter grows 
as $\gamma=1/(4Ht)$ decreases. 
Finally, the accelerated expansion stops as $\gamma \rightarrow 0$ like arriving at the critical point. 
This is an interesting scenario encompassing the exact solutions altogether. 
It is remarkable in the first place that UV effects predict the IR behavior of the theory. 
This is specific to two dimensions as the propagators of the minimally coupled scalars are scale invariant. 
The UV and IR divergences are closely related 
since $\log(P_\text{max}/P_\text{min})$-type large logarithms are expected in two dimensions.
It is very interesting to find out whether IR effects diminish the scaling dimension altogether down to nil. 

The situation is different in four dimensions, where only $\log (H/P_\text{min})$-type IR logarithms appear. 
The solutions of Fokker-Planck equations do not depend on $Q^2$ or contain all of them. 
It is a characteristic feature of the one-loop approximation which takes account of the leading IR logarithms. 
On the other hand, the exact solution of the renormalized action shows explicit $Q^2$ dependence 
indicating all loop contributions.
The picture we obtain from the exact solutions is not only more sophisticated but also very convincing. 

Apparently the effective field theory for the exactly renormalized action contains inflatonlike freedom. 
Such freedom seems necessary to describe the solution of quantum gravity 
in terms of the effective theory at the tree level. 
It is in some sense a dual description of quantum gravity. 
Remarkably, its classical behavior reproduces quantum effects. 
We wonder whether the inflaton in our Universe may be such a dual description of quantum effects. 

\section{Conclusions}
\setcounter{equation}{0}

We have investigated IR quantum effects in the two-dimensional de Sitter space 
from a solution of the exactly renormalized Liouville action. 
We work in the semiclassical region where the matter central charge $c > 25$. 
In such a region, the exact scaling dimension $\gamma$ of the cosmological constant operator
is less than $\gamma<1$. 
This is due to the screening effect of the conformal mode with a negative metric. 
The two-dimensional de Sitter space is obtained as a solution of the Liouville action. 
The solution of the renormalized action shows that two-dimensional de Sitter space is doomed. 
The Hubble parameter is no longer constant and decreases with time. 
It does so slowly at the weak coupling with large $c$ and even stops acceleration at the critical point $c=25$. 

In conclusion, we have made a strong case for the instability of two-dimensional de Sitter space. 
The exact solutions show that the negative anomalous dimension of the cosmological constant operator
makes the Hubble parameter time dependent and vanishes at a late time. 
They underscore the importance of IR logarithms which become shielding effects 
due to the negative sign of the kinetic term of the conformal mode. 
We estimate the semiclassical entropy of two-dimensional de Sitter space. 
It increases logarithmically with the Hubble radius as $\log (1/H^2(t))$ versus $1/H^2(t)$ in four dimensions. 
The cosmological constant is diffused by entropy production at the horizon. 
The conformal zero mode generates entropy in a Brownian diffusion process. 
We formulate the Fokker-Planck equation in two-dimensional quantum gravity. 
We take account of the change of the conformal zero mode distribution 
by the renormalization of the cosmological constant operator. 
In this way we can obtain very analogous equations with unitary matter systems
despite the negative metric of the conformal mode. 
Nevertheless we argue that the drift term is absent 
due to the uniqueness of the classical solution. 

We propose that the de Sitter entropy is carried by the conformal zero modes. 
In order to verify our proposal, 
we have evaluated the von Neumann entropy of the distribution functions for them. 
Their characteristics are in agreement with semiclassical estimates. 
In matter systems, it is known that the equilibrium state is stable. 
Since we have only the quantum fluctuation term, the system is diffused away with the Hubble parameter 
to vanish at a late time. 
In the matter systems, the free energy $F=E-TS$ is minimized. 
At low temperature, minimizing the energy is important. 
In a standard model, we look for the ground state with the smallest energy. 
On the other hand, we maximize the entropy as there is no energy in de Sitter space. 
It should be interesting to understand such an evolution which takes place in quantum gravity. 
This perspective may shed new light on the fine-tuning problem  
since the maximum entropy principle operates in quantum gravity \cite{Nielsen,HKAWAI}. 
In fact, the cosmological constant may turn out be such an example. 
There are many common features between two-dimensional and four-dimensional gravity 
such as the negative sign of the kinetic term of the conformal mode. 
We hope to investigate the relation between the cosmological constant 
and the generation of entropy of the superhorizon mode in four-dimensional Einstein gravity. 

\section*{Acknowledgment}
This work is supported by the National Center of Theoretical Sciences (NCTS) 
and Grant-in-Aid for Scientific Research (C) No. 16K05336. 
We thank Chong-Sun Chu, Satoshi Iso, Hikaru Kawai, Yoji Koyama and Sanjin Shin for discussions. 
Y. K. thanks  Kimyeong Lee  for discussions and his warm hospitality at KIAS.

\appendix

\section{de Sitter thermodynamics}\label{A}
\setcounter{equation}{0}

Globally, de Sitter space is a hyperboloid,  
\begin{align}
\frac{ds^2}{l^2}=-d\tau^2+\cosh^2(\tau)d\Omega_3^2 .
\label{GDS}\end{align}
The characteristic length $l$ is set by the Hubble parameter $H$ as $l=1/H$. 
It is related to  the cosmological constant $\Lambda$ as $l= \sqrt{3/\Lambda}$. 
Locally in the Poincar\'e coordinate, it corresponds to an expanding flat universe, 
\begin{align}
ds^2=-dt^2+e^{2Ht}(dr^2+r^2d\Omega_2^2)=\big(\frac{1}{-H\tau}\big)^2(-d\tau^2+d\vec{x}^2), 
\label{LPC}\end{align}
where we can obtain conformally flat parametrization.
We often work in this coordinate where $\tau$ runs from $-\infty$ to $0$. 
Let us draw an $S^2$ with the radius $a(t)r$ in this space where $a(t)=e^{Ht}$. 
The expansion velocity of this space is $Ha(t)r$. 
Since it coincides with the velocity of light at the apparent horizon, its radius is $\rho_h=1/H$.
In the static coordinate system, 
\begin{align}
\frac{ds^2}{l^2}=-V(r)dt^2+\frac{1}{V(r)}dr^2+r^2d\Omega_2^2, 
\end{align}
where $V(r)=1-r^2$. 
It becomes manifest that an observer at $r=0$ is surrounded by a cosmological horizon at $r=1$. 
The radius of the horizon is $\rho_h=1/H$ in agreement with that in the Poincar\'e coordinate. 

The presence of event horizons leads to thermodynamics \cite{Bousso}.
According to Bekenstein and Hawking, black holes possess finite temperature: 
\begin{align}
T_\text{hor}=\frac{\kappa}{2\pi}. 
\end{align}
For a Schwarzschild black hole of mass $M$, $\kappa=1/{4M}$. 
The first law of thermodynamics is
\begin{align}
\frac{1}{T_\text{hor}}=\frac{\partial S}{\partial M}. 
\end{align}
The entropy is given by the area of the event horizon, 
\begin{align}
S_\text{hor}=\frac{A}{4} .
\end{align}

We list here the relationship among different coordinates on two-dimensional de Sitter space 
or its Euclidean version $S^2$. 
The metric on $S^2$ is
\begin{align}
ds^2=\frac{1}{H^2}(d\theta^2+\sin^2 (\theta)d\varphi^2). 
\end{align}
It can be embedded into three Euclidean dimensions.
\begin{align}
z_0^2+z_1^2+z_2^2=\big(\frac{1}{H}\big)^2. 
\end{align}
We rotate it into real spacetime: $z_0\rightarrow iz_0$, 
\begin{align}
-z_0^2+z_1^2+z_2^2=\big(\frac{1}{H}\big)^2. 
\end{align}
We then consider the following coordinate: 
\begin{align}
z_0=\frac{1}{H}\sinh(Ht)+\frac{1}{2}He^{Ht}x^2,\hspace{1em}
z_2=\frac{1}{H}\cosh(Ht)-\frac{1}{2}He^{Ht}x^2,\hspace{1em}
z_1=e^{Ht}x, 
\end{align}
with the line element
\begin{align}
ds^2=-dt^2+e^{2Ht}dx^2=\big(\frac{1}{-H\tau}\big)^2(-d\tau^2 +dx^2), 
\end{align}
which covers the upper half-triangle of the Penrose diagram of the global de Sitter manifold. 
The metric on $S^2$ can be continued to the global de Sitter metric more directly $\theta=\frac{\pi}{2}+it$: 
\begin{align}
ds^2=\frac{1}{H^2}(-dt^2+\cosh^2 (t )d\varphi^2). 
\end{align}

Although we mostly work in the Poincar\'e coordinate, we may rotate it  into $S^2$
by using these relations if appropriate. 
The coordinate transformation can be done by inspection. 
For example, we claim the following term is topologically quantized after Euclidean rotation into $S^2$: 
\begin{align}
\frac{1}{8\pi}\int d\tau dx \frac{1}{(-H\tau )^2 }{2H^2}=\frac{1}{8\pi}\int d^2x \sqrt{-g}R\ 
\rightarrow\ -i\frac{1}{4\pi}\int d\theta d\omega \sin \theta  = -i. 
\end{align}

The two-dimensional Liouville gravity may be thought of a little Einstein gravity descending 
from four dimensions to $D=2+\epsilon$ dimensions: 
\begin{align}
\frac{Q^2}{4\pi}\int d^Dx\sqrt{g}(\frac{1}{\epsilon}R - H^2)
=\frac{Q^2}{4\pi}\int d^Dx\sqrt{-\hat{g}}(\frac{1}{\epsilon}e^{\frac{\epsilon}{2}\phi}{\hat{R}}-e^{\phi}H^2), 
\end{align}
where we explicitly show the constant conformal mode dependence 
\begin{align}
g_{\mu\nu}=e^{\phi}\hat{g}_{\mu\nu}. 
\end{align}
The first term with the $1/\epsilon$ pole is
\begin{align}
\frac{1}{\epsilon}\frac{Q^2}{4\pi}\int d^2x \sqrt{g} R 
&=\frac{Q^2}{4\pi}\int d^2x\sqrt{\hat{g}}(\frac{1}{\epsilon}\hat{R}
+\frac{1}{4}\phi\hat{g}^{\mu\nu}\nabla_\mu\nabla_\nu\phi 
+\frac{1}{2}\hat{g}^{\mu\nu} \nabla_\mu\phi \nabla_\nu\phi +\frac{1}{2}\phi \hat{R}) \notag\\
&=\frac{Q^2}{4\pi}\int d^2x\sqrt{\hat{g}}(\frac{1}{\epsilon}\hat{R}
+\frac{1}{4}\hat{g}^{\mu\nu} \nabla_\mu\phi \nabla_\nu\phi +\frac{1}{2}\phi \hat{R}). 
\end{align}
The leading term with the $1/\epsilon$ pole acts as the counterterm.
We thus obtain 
\begin{align}
\frac{Q^2}{4\pi}\int d^2x\sqrt{\hat{g}}(\frac{1}{4}\hat{g}^{\mu\nu} \partial_\mu\phi \partial_\nu\phi 
+\frac{1}{2}\phi \hat{R}-H^2). 
\end{align}
This is the Liouville action which is the gift of conformal anomaly.

\section{Renormalization of cosmological constant operator }\label{B}
\setcounter{equation}{0}

To renormalize the cosmological constant operator to the leading order in $1/Q^2$, 
we need to consider the quantum fluctuation of the cosmological constant operator: 
\begin{align}
\langle e^{\phi} \rangle = \langle e^{\phi_c+\tilde{\phi}} \rangle 
\sim e^{\phi_c(t)+\frac{1}{2}\langle \tilde{\phi}^2 \rangle}. 
\end{align}
Here $\phi_c(t)$ denotes a classical solution while
\begin{align}
\langle 
\tilde{\phi}^2 \rangle=-\frac{4}{Q^2}\int_{P_\text{min}}^{P_\text{max}} \frac{dP}{P}. 
\end{align}
The scalar propagator is both UV and IR divergent in two dimensions. 
We recall here that the physical momenta $P$ depend on the metric 
\begin{align}
P_{\text{max}}=\frac{p_{\text{max}}}{e^{\frac{1}{2}\tilde{\phi} (x)}}. 
\end{align}

We first consider the UV contribution in a dimensional regularization. 
We consider $D=2-\epsilon$ dimensions since the $1/\epsilon$ pole can be identified 
with the logarithmic UV divergence. 
The cosmological constant operator is
\begin{align}
e^{\frac{D}{2}\tilde{\phi}}=e^{(1-\frac{\epsilon}{2})\tilde{\phi}} .
\end{align}
We evaluate the two-point function as
\begin{align}
\frac{1}{2} \langle \tilde{\phi}^2 \rangle 
&=-\frac{2}{Q^2}\int {dp}{p^{1-\epsilon}}\frac{1}{p^2+m^2}\mu^{\epsilon}
e^{-\frac{1}{2}\epsilon\tilde{\phi}} \notag\\
&=-\frac{1}{Q^2}\Gamma \big(\frac{\epsilon}{2}\big)m^{-\epsilon}
\mu^{\epsilon}e^{-\frac{1}{2}\epsilon\tilde{\phi}} \notag\\
&\sim -\frac{2}{Q^2}\big(\frac{1}{\epsilon}+\log\frac{\mu}{m}-\frac{1}{2}\tilde{\phi}\big),  
\label{2PTFN}\end{align} 
where $\mu$ is the renormalization scale. 
We set $\mu=m$ for simplicity. 
One of the difficulties of renormalizing the operators in quantum gravity is that both the propagators 
and the interaction vertices depend on the metric. 

To carry over this renormalization process to all orders, 
we employ a renormalization group \cite{KKN1,KKN}. 
Let us recall our dimensional regularized Lagrangian
\begin{align}
\int d^Dx \frac{Q^2}{8\pi}\big\{\frac{1}{2}\eta^{\mu\nu}e^{-\frac{\epsilon}{2}\phi} 
\frac{\partial}{ \partial x^\mu}\phi\frac{\partial}{\partial x^\nu}\phi
-2H^2e^{(1-\frac{\epsilon}{ 2})\phi}\big\}. 
\end{align}
It is always a good idea to canonically normalize the kinetic term 
by the change of field variable $e^{-\frac{\epsilon}{4}\phi} =1-\frac{\epsilon}{4}\psi$, 
\begin{align}
\int d^D x \frac{Q^2}{8\pi}\big\{\frac{1}{2}\eta^{\mu\nu}
\frac{\partial}{\partial x^\mu}\psi\frac{\partial}{\partial x^\nu}\psi 
-2H^2(1-\frac{\epsilon}{4}\psi)^{-\frac{4}{\epsilon}(1-\frac{\epsilon}{2})}\big\}.
\end{align}
Now the two-point function is just (\ref{2PTFN}) without $\psi$ field dependence: 
\begin{align}
\frac{1}{2} \langle \tilde{\psi}^2 \rangle 
\sim-\frac{2}{Q^2}(\frac{1}{\epsilon}+\log\frac{\mu}{ m}) .
\label{2PTFNM}\end{align} 

The advantage of this approach is that we need not worry about the interaction vertices in the kinetic term.
Let us investigate the quantum correction to the cosmological constant operator next: 
\begin{align}
(1-\frac{\epsilon}{4}\psi)^{-\frac{4}{\epsilon}(1-\frac{\epsilon}{2})}
&=\exp\big\{-\frac{4}{\epsilon}(1-\frac{\epsilon}{2})\log(1-\frac{\epsilon}{4}\psi)\big\}\notag\\
&=\exp\big\{(1-\frac{\epsilon}{2})(\psi + \frac{\epsilon}{4}\frac{1}{2}\psi^2+
 \frac{\epsilon^2}{4^2}\frac{1}{3}\psi^3+\cdots\big\}. 
\end{align}
The one-loop quantum corrections start with the $e^{\psi}$ part of the operator
\begin{align}
\langle\exp \big\{(1-\frac{\epsilon}{2})\psi\big\}\rangle 
&\sim 1+(1-\frac{\epsilon}{2})\bar{\psi}+\frac{1}{2}(1-{\epsilon})\langle\psi^2\rangle
+\frac{1}{2}(1-\frac{3\epsilon}{2})\langle\psi^2\rangle\bar{\psi} \notag\\
&=\big\{1-(1-\epsilon)\frac{2}{Q^2}\frac{1}{\epsilon}\big\}\big\{1+(1-\frac{\epsilon}{2})\bar\psi\big\}. 
\end{align}
There are additional contributions, 
\begin{align}
\langle\frac{\epsilon}{4}\frac{1}{2}\psi^2\rangle
+\langle\frac{\epsilon}{4}\frac{1}{2}\psi^3\rangle 
=-\frac{1}{2}\frac{1}{Q^2}-\frac{3}{2}\frac{1}{Q^2}\bar{\psi} .
\end{align}
By a multiplicative renormalization by $Z$, we obtain the renormalized operator at the one-loop level: 
\begin{align}
&Z\big\{1-(1-\epsilon)\frac{2}{Q^2}\frac{1}{\epsilon}-\frac{1}{2}\frac{1}{Q^2}\big\}
\big\{1+(1-\frac{1}{Q^2})(1-\frac{\epsilon}{2})\bar{\psi}\big\} \notag\\
=&1+(1-\frac{1}{Q^2})(1-\frac{\epsilon}{2})\bar{\psi} 
\sim \exp \big\{(1-\frac{1}{Q^2})(1-\frac{\epsilon}{2})\bar{\psi}\big\}. 
\end{align}

Let us introduce a trick to determine the UV divergence of a generic operator.
We consider the following integral weight: 
\begin{align}
\sqrt{\frac{\epsilon Q^2}{8\pi}}\int d\psi e^{-\frac{\epsilon Q^2}{8}\psi^2}, 
\label{ITMS}\end{align}
such that the average of the two-point function produces its $1/\epsilon$ pole in $2+\epsilon$ dimensions: 
\begin{align}
\sqrt{\frac{\epsilon Q^2}{8\pi}}\int d\psi e^{-\frac{\epsilon Q^2}{8}\psi^2}\psi^2= \frac{4}{\epsilon Q^2}. 
\end{align}
We split the field such that $\psi\rightarrow \psi_c+\psi$ 
and take the average over the $\psi$ field with this measure first. 
We can determine its UV divergences this way.

A generic proof is
\begin{align}
\langle\frac{1}{2l!}(\psi_c+\psi)^{2l}\rangle_-
&=\langle\sum\frac{1}{2m!} \psi_c^{2m}\frac{1}{2n!}\psi^{2n}\rangle_-
\notag\\
&=\sum\frac{1}{2m!}\psi_c^{2m}\frac{1}{n!}\big(\frac{1}{2}\langle\psi^2\rangle_-\big)^n, 
\end{align}
where the average denoted by $\langle\psi^{2n}\rangle_-$ is with respect to the weight (\ref{ITMS}). 
It is also clear that 
\begin{align}
\exp\big(-\frac{2}{\epsilon Q^2}\frac{\partial^2}{\partial \psi ^2}\big)F(\psi)
\label{NBO}\end{align}
is the finite operator. 
This is because 
\begin{align}
&\exp\big(-\frac{2}{\epsilon Q^2}\frac{\partial^2}{\partial \psi_c ^2}\big) 
\frac{1}{2l!}\langle(\psi_c+\psi)^{2l}\rangle_- \notag\\
=&\exp\big(-\frac{2}{\epsilon Q^2}\frac{\partial^2}{\partial \psi_c ^2}\big)
\sum\frac{1}{2m!}\psi_c^{2m}\frac{1}{n!}(\frac{1}{2}\langle\psi^2\rangle_-)^n \notag\\
=&\sum \frac{1}{m!}(-\frac{2}{\epsilon Q^2})^m \frac{1}{n!}(\frac{1}{2}\langle\psi^2\rangle_-)^n\notag\\
=&\frac{1}{l!}\big\{(-\frac{2}{\epsilon Q^2})+ (\frac{2}{\epsilon Q^2})\big\}^l=0. 
\end{align}

Let us introduce the renormalization scale $\mu$ according to its canonical dimension. 
In doing so, we have introduced an arbitrary scale $\mu$ 
in the bare inverse coupling $Q_B^2=Q^2\mu^{-\epsilon}$. 
Since the bare coupling cannot depend on how to decompose it, we conclude that $Q^2 \sim \mu^{\epsilon}$. 
By demanding $\mu$ independence on the bare operator (\ref{NBO}), 
we can derive a renormalization group equation for the renormalized operator 
\begin{align}
\mu\frac{\partial}{\partial \mu}F = -\frac{2}{Q^2}\frac{\partial^2}{\partial \psi ^2}F .
\label{DFEQ}\end{align}
This equation does not depend on the sign of $\epsilon$ .
The operator $F$ diffuses at long distances. 

In fact, the solution of this diffusion equation coincides with the finite cosmological constant operator 
constructed by the integral measure (\ref{ITMS}). 
It is the diffusion kernel where diffusion time is identified 
with $1/\epsilon \sim- \log \mu$ in $2+\epsilon$ dimensions. 
In the two-dimensional limit, it can be exactly calculated as follows 
\begin{align}
&\sqrt{\frac{\epsilon Q^2}{8\pi}}\int d\psi e^{-\frac{\epsilon Q^2}{8}\psi^2}
e^{\frac{4}{\epsilon}(1+\frac{\epsilon}{2})\log\{1+\frac{\epsilon}{4}(\psi_c+\psi)\}} \notag\\
=&\sqrt{\frac{\epsilon Q^2}{8\pi}}\int d\psi e^{-\frac{\epsilon Q^2}{8}(\psi-\psi_c)^2}
e^{\frac{4}{\epsilon}(1+\frac{\epsilon}{2})\log(1+\frac{\epsilon}{4}\psi)} \notag\\
=&\sqrt{\frac{\epsilon Q^2}{8\pi}}\frac{4}{\epsilon}\int d\rho 
e^{-\frac{2 Q^2}{\epsilon}(\rho-\frac{\epsilon}{4}\psi_c)^2}
e^{\frac{4}{\epsilon}(1+\frac{\epsilon}{2})\log(1+\rho)}
\sim e^{Q^2\rho_0\psi_c} .
\end{align}
In the $\epsilon \rightarrow 0$ limit, $\rho_0$ is determined by the saddle-point approximation
which leads (\ref{saddle}), and the scaling dimension is determined as
\begin{align}
Q^2\rho_0=\gamma .
\end{align}
The renormalized cosmological constant operator is determined to be $e^{\gamma\phi}$
in agreement with the conformal invariance approach (\ref{anmd}). 
The advantage of this approach is that 
it demonstrates how the original cosmological constant operator at short distances 
evolves toward the renormalized form at long distances due to quantum effects \cite{KKN}. 

The Fokker-Planck equation is also a diffusion equation, 
\begin{align}
\dot{\rho}=2H\frac{1}{Q^2}\frac{\partial^2}{\partial \phi^2} \rho, 
\end{align}
with the  identification $Ht=\mu^{-\epsilon}/\epsilon \sim -\log \mu$ to relate it to (\ref{DFEQ}). 
Let us construct the diffusion kernel in the conjugate variables to $\phi$: 
\begin{align}
K=\frac{1}{\sqrt{2}}e^{-t2H\frac{1}{Q^2}p^2},\hspace{1em}\frac{\partial}{\partial t}K=-2H\frac{1}{Q^2}p^2 K. 
\end{align}
The solution in the dual variables is given by
\begin{align}
\rho{(t,p)}=K(t,p)\rho(p). 
\end{align}
After the Fourier transformation, we obtain
\begin{align}
\rho_t
&=\int d\phi ' K(t,\phi-\phi')\rho(\phi') \notag\\
&=\int d \phi' \frac{Q}{\sqrt{8Ht\pi}}e^{-\frac{Q^2(\phi-\phi ')^2}{8Ht}}
\frac{Q}{\sqrt{2\pi}} e^{-\frac{Q^2}{2}\phi '^2}\notag\\
&=\sqrt\frac{ Q^2}{ 2\pi(1+4H t)}e^{-\frac{Q^2}{ 2(1+4H t)}\phi^2} .
\end{align}

There is an alternative method to impose the conformal invariance on the bare operator 
as was mentioned before. 
In this approach, we construct the bare operator which is invariant 
under $\phi_c\rightarrow\phi_c-\varphi,\ \tilde{\phi}\rightarrow\tilde{\phi}+\varphi$ \cite{DK,FD}. 

The one-loop short-distance divergence is evaluated in a dimensional regularization (\ref{2PTFN}): 
\begin{align}
\langle e^{\gamma\phi} \rangle = \exp \big(\frac{\gamma^2}{2}\langle\phi^2\rangle\big)
= \exp\big\{-\frac{2\gamma^2}{Q^2}
\big(\frac{1}{\epsilon}-\frac{1}{2}\tilde{\phi}\big)\big\}. 
\label{DDK}\end{align}
The bare operator is constructed by subtracting the UV-cutoff-dependent part, 
\begin{align}
e^{\phi_c}e^{\gamma\phi}Z,\hspace{1em}
Z=\exp\big(\frac{2\gamma^2}{Q^2}\frac{1}{\epsilon}\big). 
\end{align}
Under the above transformation, the bare operator changes as
\begin{align}
e^{\gamma\phi}Z(\phi) \rightarrow e^{\gamma\varphi+\frac{\gamma^2}{Q^2}\varphi}. 
\end{align}
We find the condition
\begin{align}
\gamma+ \frac{\gamma^2}{Q^2}=1. 
\end{align}
One-loop computation is sufficient to perform the  exact renormalization 
as the self-consistent solution is obtained.
By solving this equation, the scaling dimension of the cosmological constant operator is determined to all orders.
It matches with the leading-order renormalization process we carried out here: 
\begin{align}
\gamma=\frac{2}{1+\sqrt{1+\frac{4}{Q^2}}}= 1-\frac{1}{Q^2} + 2\big(\frac{1}{Q^2}\big)^2+\cdots.
\end{align}

\section{Uniqueness of the duality}\label{C}
\setcounter{equation}{0}

Martinec and Moore considered the following action, 
which contains not only timelike but also spacelike Liouville fields $\phi$ and $\varphi$ respectively \cite{MM}: 
\begin{align}
\int \sqrt{-\hat{g}}d^2x
\big\{&\frac{Q^2}{16\pi}(\hat{g}^{\mu\nu}\partial_\mu\phi\partial_\nu\phi 
+2\phi\hat{R}-\frac{4H^2}{\gamma^2} e^{\gamma\phi}) \notag\\
&-\frac{q^2}{16\pi}(\hat{g}^{\mu\nu}\partial_\mu\varphi\partial_\nu\varphi
+2\varphi\hat{R})\big\}. 
\end{align}
where the total central charge must vanish: $6Q^2-6q^2-c_\text{matt}+24=0$. 
This is the Liouville gravity with the cosmological constant for the $\phi$ field 
while the $\varphi$ field is free in the conformal gauge. 
The scaling dimension of the cosmological constant operator $\gamma$ is determined by $Q^2$. 

The equations of motion with respect to $\phi$ and $\varphi$ are 
\begin{align}
\nabla_0^2\phi=\frac{1}{\gamma}H^2e^{\gamma\phi},\hspace{1em}
\nabla_0^2\varphi=0,  
\label{CEQ1}\end{align}
whose solutions are given by 
\begin{align}
\phi_c= -\frac{2}{\gamma}\log(-H\tau),\hspace{1em}\varphi_c=0. 
\label{Sol1}\end{align}

The timelike and spacelike Liouville fields can be mixed by a hyperbolic rotation: 
\begin{align}
Q\phi=Q\tilde{\phi}c -q\tilde{\varphi}s,\hspace{1em}
q\varphi=q\tilde{\varphi}c -Q\tilde{\phi}s, 
\end{align}
\begin{align}
\tilde{Q}=Qc + qs,\hspace{1em}
\tilde{q}=qc + Qs, 
\end{align}
where $c=\cosh (\lambda),\ s=\sinh(\lambda)$. 
This process produces a class of formally equivalent inflaton models 
which are interesting testing grounds for our understanding about inflation: 
\begin{align}
\int \sqrt{-\hat{g}}d^2x 
\big\{&\frac{Q^2}{16\pi}(\hat{g}^{\mu\nu}\partial_\mu\tilde\phi\partial_\nu\tilde{\phi}
+2\frac{\tilde{Q}}{Q}\tilde{\phi}\hat{R}
-\frac{4H^2}{\gamma^2} e^{{\gamma}(\tilde{\phi}c -\tilde{\varphi}\frac{s}{Q})}) \notag\\
&-\frac{1}{16\pi}(\hat{g}^{\mu\nu}\partial_\mu\tilde{\varphi}\partial_\nu\tilde{\varphi}
+2\tilde{q}\tilde{\varphi}\hat{R})\big\}, 
\label{MM-L}\end{align}
where $\tilde{\varphi}$ is normalized as $q\tilde{\varphi}\to\tilde{\varphi}$ 
and can be identified as an inflaton. 

The equations of motion with respect to the conformal mode and the inflaton are 
\begin{align}
\nabla_0^2\gamma\tilde{\phi}
= 2H^2c e^{{\gamma}(\tilde{\phi}c -\tilde{\varphi}\frac{s}{Q})},\hspace{1em}
\nabla_0^2\gamma\tilde{\varphi}
= 2H^2Qs e^{{\gamma}(\tilde{\phi}c -\tilde{\varphi}\frac{s}{Q})}. 
\label{CEQ2}\end{align}
These equations are solved as 
\begin{align}
\tilde{\phi}_c=-\frac{2c}{\gamma}\log(-H\tau),\hspace{1em}
\tilde{\varphi}_c= Q\frac{s}{c}\tilde{\phi}_c. 
\label{Sol2}\end{align}

We still need to satisfy the equation of motion 
with respect to the traceless mode of the metric $h^{\mu\nu}$: 
\begin{align}
\frac{Q^2}{8\pi}(\nabla_\mu\tilde{\phi}\nabla_\nu\tilde{\phi}
-2\frac{\tilde{Q}}{Q}\nabla_\mu\nabla_\nu\tilde{\phi})
=\frac{1}{8\pi}(\nabla_\mu\tilde{\varphi}\nabla_\nu\tilde{\varphi}-2\tilde{q}\nabla_\mu\nabla_\nu\tilde{\varphi})
+\nabla_\mu\chi\nabla_\nu\chi, 
\label{HEQ2}\end{align}
where $\chi$ denotes a free scalar field.\footnote{
Strictly speaking, both sides of (\ref{HEQ2}) should be made traceless as in (\ref{HEQ}). 
However, such a process is not necessary if we are concerned only with homogeneous background fields. }
In terms of the old unrotated variables $\phi$ and $\varphi$, 
this equation can be simplified as 
\begin{align}
\frac{Q^2}{8\pi}(\nabla_\mu \phi\nabla_\nu \phi-2\nabla_\mu\nabla_\nu \phi)=0, 
\label{HEQ1}\end{align}
where we have substituted the trivial solutions $\varphi_c=\chi_c=0$ on the right-hand side. 
It should be noted that if $\gamma<1$, the left-hand side no longer vanishes 
for the solution $\phi_c=-\frac{2}{\gamma}\log (-H\tau)$.  
We need to identify something to fill this gap. 

Our proposal is to interpret an inflaton as a quantum degree of freedom and assign this role to it. 
Specifically, we introduce a minimally coupled inflaton $f$ into the action as follows: 
\begin{align}
\int d^2 x \frac{Q^2}{8\pi}\sqrt{-\hat{g}}
\big\{-\frac{1-\gamma}{2}\hat{g}^{\mu\nu}\partial_\mu f \partial_\nu f
+\frac{1}{2}(\hat{g}^{\mu\nu}\partial_\mu \phi \partial_\nu \phi +2 \hat{R}\phi)
-\frac{2H^2}{\gamma}e^{\phi-(1-\gamma)f}\big\}, 
\end{align}
where $f$ has no $\hat{R}f$ term in contrast to $\tilde{\varphi}$ in (\ref{MM-L}). 

The equations of motion with respect to the inflaton $f$ and the conformal mode $\phi$ are 
\begin{align}
\nabla_0^2\gamma f = 2H^2e^{\phi-(1-\gamma)f},\hspace{1em}
\nabla_0^2\gamma \phi = 2H^2 e^{\phi-(1-\gamma)f}, 
\label{CEQ3}\end{align}
whose solutions are given by 
\begin{align}
f=\phi= -\frac{2}{\gamma}\log(-H\tau). 
\label{Sol3}\end{align}
These solutions satisfy the equation of motion with respect to $h^{\mu\nu}$: 
\begin{align}
\frac{Q^2}{8\pi}(\nabla_\mu{\phi}\nabla_\nu \phi-2\nabla_\mu\nabla_\nu \phi)
=\frac{Q^2}{8\pi}(1-\gamma)\nabla_\mu f\nabla_\nu f. 
\label{HEQ3}\end{align}
We can identify $f$ as $\phi$ by the use of the equations of motion. 
We regard this semiclassical theory as a dual description of quantum Liouville gravity. 
We do not touch a spacelike Liouville term even if it is present in the Lagrangian. 
The presence of such a term can be felt only through $Q^2$. 

The physical properties of this inflationary universe are studied in Section 2.
A class of two-dimensional quantum gravity models which are related by the change of variables 
represents a unique model in our effective Lagrangian approach. 
We believe this fact supports our interpretation that 
an inflaton emerges as a quantum effect in quantum gravity. 

\section{No drift force for conformal mode}\label{D}
\setcounter{equation}{0}

In de Sitter spaces for superhorizon modes, the effective viscosity might become large 
and the $\phi$ field moves with a velocity proportional to the potential force: 
\begin{align}
\phi\rightarrow \phi +\frac{2}{Q^2} \frac{\partial}{\partial \phi}V(\phi) \log a = \phi+ 2(e^{\phi}-1) Ht. 
\end{align}
There is no suppression factor by $1/Q^2$ here since it is a tree effect. 
They cancel between the propagator and the vertex. 
We thus obtain an analogous equation to that of the inflaton in inflation theory: 
\begin{align}
\dot{\phi}=2H(e^{\phi}-1) .
\label{RUCM}\end{align}
This equation is obtained diagrammatically 
but it must follow from the equation of motion of the conformal field $\phi$. 

The equation of motion for the quantum field $\phi$ is 
\begin{align}
\frac{\partial^2}{\partial \tau^2}{\phi}-\frac{\partial^2}{\partial x^2} \phi 
-2H^2e^{\gamma\phi_c}(e^{\phi}-1)=0 .
\label{EQQF2}\end{align}
Let us consider $\check{\phi}=\gamma\phi_c+\phi$, where $\phi_c$ is the classical solution 
which describes de Sitter space. 
We find that $\check{\phi}$ satisfies the identical equation with $\gamma\phi_c$, 
\begin{align}
\frac{\partial^2}{\partial \tau^2}{\check{\phi}}-\frac{\partial^2}{\partial x^2} \check{\phi} 
-2H^2e^{\check{\phi}}=0. 
\label{EQQF1}\end{align}
Both $\gamma\phi_c$ and $\check{\phi}$ are the solution of the Liouville theory.
As pointed out before, we have separated the original field $\phi_0=\gamma\phi_c+\phi$. 
We can consider the transformation $\gamma\phi_c \rightarrow \gamma\phi_c+\varphi$ 
and $\phi \rightarrow \phi-\varphi$, 
which leaves the theory invariant as long as $\varphi$ represents local fluctuations. 

However, it is doubtful that there are two different solutions. 
In fact, $\check{\phi}$ is the solution only when the second derivative with respect to time can be neglected, 
namely near the origin of the field space $\check{\phi} \sim 0$. 
From the equations of $\phi_c$ and $\phi$, 
\begin{align}
\gamma\phi_c=2Ht,\hspace{1em}\dot{\phi}=2H(e^{\phi}-1), 
\end{align}
we obtain 
\begin{align}
e^{2Ht}H\dot{\check{\phi}}=2H^2e^{\phi+2Ht}=2H^2e^{\check{\phi}}, 
\end{align}
while
\begin{align}
e^{2Ht}\ddot{\check{\phi}}=(2H)^2e^{\phi+{2Ht}}(e^{\phi}-1)=2H^2e^{\check{\phi}}(e^{\phi}-1) .
\end{align}
In other words $\check{\phi}$ is not the solution in other regions. 
We conclude that there is a unique de Sitter solution $\gamma\phi_c$ in this model. 
The free-field solution without the potential is
\begin{align}
\frac{1}{\sqrt{2p}}e^{-ip\tau+i\vec{p}\cdot\vec{x}}. 
\end{align}

We focus on the superhorizon mode, 
\begin{align}
\phi_0(x)=\sqrt{\frac{8\pi}{Q^2}}\int \frac{d\vec{p}}{2\pi}\theta (Ha(t)-p)
\big(a_{\vec{p}}\frac{1}{\sqrt{2p}}e^{i\vec{p}\cdot\vec{x}}
+a_{\vec{p}}^\dagger\frac{1}{\sqrt{2p}}e^{-i\vec{p}\cdot\vec{x}}\big), 
\end{align}
where $[a_{\vec{p}},a_{\vec{p}'}^\dagger]=-2\pi\delta(\vec{p}-\vec{p}')$. 
Since plane waves become constant in time, 
the time dependence is caused by the step function which restricts physical momenta $P<H$. 

The Yang-Feldman-type solution is
\begin{align}
\phi (x)=\phi_0 (x)+i\int d\tau'\int d\vec{x}' G_R(x,x') 2H^2e^{\gamma\phi_c}(e^\phi-1)(x'). 
\end{align}
The retarded propagator may be approximated for the superhorizon mode: 
\begin{align}
G_R(x,x')&\sim\theta (t-t')\int \frac{d\vec{p}}{2\pi}-i(\tau-\tau')e^{i\vec{p}\cdot(\vec{x}-\vec{x}')} \notag\\
&=-i\frac{1}{H}\theta (t-t')\delta(\vec{x}-\vec{x}')\big\{\frac{1}{a(t')}-\frac{1}{a(t)}\big\}. 
\end{align}
We thus obtain 
\begin{align}
\phi (x) = \phi_0 (x) +2H\int^t dt' (e^{\phi(t', \vec{x})}-1) .
\label{PreLang}\end{align}
By  differentiating (\ref{PreLang}), we obtain the Langevin equation: 
\begin{align}
\dot{\phi}(x) = \dot{\phi}_0 (x) + 2H (e^{\phi (x)}-1),\hspace{1em}
\langle\dot{\phi}_0 (t,\vec{x})\dot{\phi}_0 (t',\vec{x})\rangle=-\frac{4}{Q^2}H\delta (t-t'). 
\end{align}
However, we believe there are no drift force effects in Liouville gravity 
since there is no acceptable solution except $\gamma\phi_c$. 
So we are left with random noise effects only: 
\begin{align}
\dot{\phi}(x) = \dot{\phi}_0 (x),\hspace{1em}
\langle\dot{\phi}_0 (t,\vec{x})\dot{\phi}_0 (t',\vec{x})\rangle=-\frac{4}{Q^2}H\delta (t-t'). 
\end{align}

\end{document}